\documentclass[trackchanges]{aastex7}
\usepackage{booktabs}

\usepackage{amsmath}
\newcommand{\sw}[1]{\texttt{#1}}

\usepackage{subfigure}

\newcommand{\swift}[1]{\textit{Neil Gehrels Swift} #1}

\usepackage{tikz}

\begin{document}

\title{SN\,2024afav: A Superluminous Supernova with Multiple Light Curve Bumps and Spectroscopic Signatures of Circumstellar Interaction}

\author[orcid=0000-0003-0871-4641]{Harsh Kumar}
\affiliation{Center for Astrophysics \textbar{} Harvard \& Smithsonian, 60 Garden Street, Cambridge, MA 02138-1516, USA}
\affiliation{The NSF AI Institute for Artificial Intelligence and Fundamental Interactions, USA}
\email[show]{harsh.kumar@cfa.harvard.edu}

\author[orcid=0000-0003-0526-2248]{Peter K.~Blanchard}
\affiliation{Center for Astrophysics \textbar{} Harvard \& Smithsonian, 60 Garden Street, Cambridge, MA 02138-1516, USA}
\affiliation{The NSF AI Institute for Artificial Intelligence and Fundamental Interactions, USA}
\email[]{}

\author[orcid=0000-0002-9392-9681]{Edo Berger}
\affiliation{Center for Astrophysics \textbar{} Harvard \& Smithsonian, 60 Garden Street, Cambridge, MA 02138-1516, USA}
\affiliation{The NSF AI Institute for Artificial Intelligence and Fundamental Interactions, USA}
\email[]{eberger@cfa.harvard.edu}

\author[orcid=0000-0002-2866-6416]{Wasundara Athukoralalage}
\affiliation{Center for Astrophysics \textbar{} Harvard \& Smithsonian, 60 Garden Street, Cambridge, MA 02138-1516, USA}
\email[]{}

\author[orcid=0000-0002-1125-9187]{Daichi Hiramatsu}
\affiliation{Department of Astronomy, University of Florida, 211 Bryant Space Science Center, Gainesville, FL 32611-2055, USA}
\affiliation{Center for Astrophysics \textbar{} Harvard \& Smithsonian, 60 Garden Street, Cambridge, MA 02138-1516, USA}
\affiliation{The NSF AI Institute for Artificial Intelligence and Fundamental Interactions, USA}
\email[]{}

\author[0000-0001-6395-6702]{Sebastian Gomez}
\affiliation{Department of Astronomy, The University of Texas at Austin, 2515 Speedway, Stop C1400, Austin, TX 78712, USA}
\email[]{}

\author[0000-0002-1895-6639]{Moira Andrews} 
\affiliation{Las Cumbres Observatory, 6740 Cortona Drive, Suite 102, Goleta, CA 93117-5575, USA}
\affiliation{Department of Physics, University of California, Santa Barbara, CA 93106-9530, USA}
\email[]{}

\author[0000-0002-4924-444X]{K. Azalee Bostroem}  
\affiliation{Steward Observatory, University of Arizona, 933 North Cherry Avenue, Tucson, AZ 85721-0065, USA}
\affiliation{LSST-DA Catalyst Fellow}
\email[]{}

\author[0000-0003-4914-5625]{Joseph R. Farah}   
\affiliation{Las Cumbres Observatory, 6740 Cortona Drive, Suite 102, Goleta, CA 93117-5575, USA} 
\affiliation{Department of Physics, University of California, Santa Barbara, CA 93106-9530, USA}
\email[]{}

\author[0000-0003-4253-656X]{D. Andrew Howell}   
\affiliation{Las Cumbres Observatory, 6740 Cortona Drive, Suite 102, Goleta, CA 93117-5575, USA} 
\affiliation{Department of Physics, University of California, Santa Barbara, CA 93106-9530, USA}
\email[]{}

\author[0000-0001-5807-7893]{Curtis McCully} 
\affiliation{Las Cumbres Observatory, 6740 Cortona Drive, Suite 102, Goleta, CA 93117-5575, USA}
\email[]{}

\begin{abstract}
We present a comprehensive optical and near-infrared spectroscopic study of SN\,2024afav -- a hydrogen-poor superluminous supernova (SLSN-I) that peaks at $\approx -20.7$ mag and exhibits an unusual multi-bumped light curve. Our spectroscopic observations, spanning phases of $-14$ to $+160$ d, reveal several unusual features: (i) a narrow ($1,800$ km s$^{-1}$) and blueshifted ($11,000$ km s$^{-1}$) absorption from H$\alpha$ starting at $+20$ d; (ii) persistent optical and NIR \ion{He}{1} lines at all available phases, showing double absorption structure in NIR spectra at $+23$ d, with a high velocity component at a similar velocity to H$\alpha$; (iii) early appearance of nebular [\ion{O}{3}] emission starting at $\approx +50$ d; and (iv) strong [\ion{O}{2}] + [\ion{Ca}{2}] 7300 \AA\ emission complex starting at $\approx +110$ d. These unusual features, and their onset at the time of the light curve bumps, provide compelling evidence of circumstellar interaction between the SN ejecta and a nearby hydrogen-rich shell, as well as the presence of helium in both the outer layers of the progenitor star and in the circumstellar medium. A comparison of SN\,2024afav to other SLSNe-I showing bumpy light curves and similar spectral properties (PTF10hgi, SN\,2017egm, SN\,2019hge), points to a rare sub-group of SLSNe-I in which CSM interaction provides an important modulation to the energy input.


\end{abstract}


\section{Introduction}\label{sec:intro}

Hydrogen-poor superluminous events are one of the brightest stellar explosion events in the Universe, with an absolute magnitude M $\lesssim -21$ and total radiated energy exceeding 10$^{50}$-10$^{-51}$ ergs~\citep{2011Natur.474..487Q, 2011ApJ...743..114C, 2012Sci...337..927G, 2015MNRAS.452.3869N, 2017MNRAS.468.4642I, 2017hsn..book..431H, 2018ApJ...852...81L, 2021A&G....62.5.34N, 2024arXiv240712302M}. Typically discovered in low-metalicity dwarf galaxies~\citep{2014ApJ...787..138L, 2016ApJ...830...13P, 2017MNRAS.470.3566C, 2024ApJ...961..169H}, these events exhibit characteristic \ion{O}{2} and \ion{C}{2} absorption features in their early-time spectra, indicating high temperatures, and show no signs of hydrogen (H) lines~\citep{2016MNRAS.458.3455M, 2016ApJ...826...39N, 2018ApJ...855....2Q, 2019MNRAS.487.2215A, 2019ApJ...882..102G, 2019ARA&A..57..305G, 2025MNRAS.541.2674A}. Recent large-scale sample studies of these events by ~\citet{2024MNRAS.535..471G}, ~\citet{2023ApJ...943...41C}, and ~\citet{2025MNRAS.541.2674A} have revealed a significant diversity in their light curve behavior and spectral properties, challenging the existing progenitor models.

The powering mechanism of SLSNe-I has been debated over the last few years. Several models of SLSNe-I powering mechanism have been proposed such as a highly spinning, magnetized neutron star (magnetar) spin-down model~\citep{2010ApJ...717..245K, 2010ApJ...719L.204W, 2014ApJ...787..138L, 2015MNRAS.454.3311M, 2016ApJ...830...13P, 2017MNRAS.470.3566C, 2017ApJ...850...55N, 2022ApJ...933...14H, 2024MNRAS.535..471G}, interaction between dense ejecta and circumstellar medium (CSM)~\citep{2016ApJ...829...17S, 2017ApJ...851L..14W, 2018NatAs...2..887L, 2020ApJ...902L...8Y, 2021MNRAS.502.2120F}, pulsational/pair-instability supernova~\citep{2009Natur.462..624G, 2015MNRAS.454.4357K, 2022A&A...666A..30P,2023NatAs...7..779L} or fallback accretion~\citep{2013ApJ...772...30D, 2018ApJ...867..113M}. Among these, magnetar models are potentially the only ones that can consistently explain the peak luminosities of SLSNe based on large-scale light curve fits, as shown in several recent studies~\citep{2021ApJ...921..180H, 2017ApJ...850...55N, 2024MNRAS.535..471G}.

While the magnetar spin-down model provides a good fit for the majority of the events, some SLSNe-I exhibit bumpy light curves that deviate significantly from the smooth spin-down curve of the magnetar model. For example, a well-studied SN\,2017egm shows a chaotic light curve which has been explained via a complex CSM interaction model~\citep{2023ApJ...949...23Z, 2023NatAs...7..779L}. Similar to SN\,2017egm, some other SLSNe-I favour CSM interaction models over magnetar~\citep{2016ApJ...829...17S, 2017ApJ...851L..14W, 2018NatAs...2..887L, 2020ApJ...902L...8Y, 2021MNRAS.502.2120F}.

At least 30\%, if not more, SLSNe-I exhibit multiple bumps in their post-peak light curves, unexplained by the favorite magnetar model~\citep{2022ApJ...933...14H}. A few such examples are well-studied events, such as PTF12dam~\citep{2017ApJ...835..266T}, iPTF13ehe~\citep{2015ApJ...814..108Y}, SN\,2015bn~\citep{2018ApJ...866L..24N}, 2017egm~\citep{2023ApJ...949...23Z}, PTF10hgi~\citep{2020ApJ...902L...8Y}, SN\,2019hge~\citep{2020ApJ...902L...8Y}, and SN\,2019unb~\citep{2021MNRAS.508.4342P}, among others. The origin of these bumps has been a long-standing mystery. For some of the events where bumps appear to be somewhat periodic, ~\citet{2024ApJ...970L..42Z, 2025ApJ...985..172Z, 2025arXiv250908051F} used a magnetar precession-based model to explain their origin. Bumps in SN\,2017egm, SN\,2017gci and other similar events were explained through CSM interaction models~\citep{2018ApJ...853...57B, 2021MNRAS.502.2120F, 2023ApJ...949...23Z}. While these models attempt to explain the early-time light curve bumpy behavior, the majority of these events lack sufficient light curve coverage to test these models over a longer period, where well-followed events like SN\,2015bn clearly show non-periodic behavior~\citep{2018ApJ...866L..24N}, suggesting a need for detailed follow-up of such events and spectroscopic study to unveil the mystery of the origin of bumps in SLSNe-I.

A handful of these bumpy events have shown very unusual spectroscopic signatures that deviate from the SLSNe-I spectra template. PTF10hgi and other SLSNe-I present in ~\citet{2020ApJ...902L...8Y} are illustrative examples of such a sample. These objects show signs of helium and, in some cases, hydrogen as well, in their spectra. \citet{2020ApJ...902L...8Y} study associates the presence of \ion{He}{1} with bumps, suggesting that CSM interaction can provide a natural mechanism to excite helium and produce bumps in the light curves of these events. Similarly, in SN\,2017egm the presence of helium and [\ion{O}{3}] is associated with a helium-rich CSM interaction with ejecta that is mainly dominated by oxygen, producing both features~\citep{2018ApJ...853...57B, 2023ApJ...949...23Z}. These events suggest that there exists a subclass of events where CSM interaction plays a crucial role in shaping the light curve and spectral features. These events may have a different progenitor channel or chaotic mass-loss history than those of typical SLSNe-I. A deeper understanding of the correlation between light curve bumps, the presence of helium, and other unusual spectral features in the spectra is crucial for understanding the progenitor and explosion physics of these events.

We present SN\,2024afav---a M $\approx -20.7$ luminosity SLSN-I with multiple bumps in the light curve. In the first paper of a two-part paper series, we present a precessing magnetar model that can explain the first few bumps~\citet{2025arXiv250908051F}. The study suggested that the presence of late-time bumps could be explained by possible interaction with the CSM. Here, we present a detailed spectroscopic study of SN\,2024afav to investigate the presence of CSM interaction and its implications on the light curve. Our extensive spectroscopic campaign reveals that SN\,2024afav exhibits rare features, including the presence of early [O III] lines, the detection of helium, the emergence of a narrow hydrogen absorption line, and the appearance of late-time [O II] lines. We provide a detailed discussion on light curve morphology, temperature evolution, spectral evolution, and a possible breakthrough in the detection of CSM interaction in SLSNe-I. This study represents a significant step towards a more detailed and systematic analysis of a large-scale sample of SLSNe-I spectra with bumpy light curves. The paper is outlined as follows: In \S\ref{sec:obs} we describe the photometric \& bolometric light curve, spectroscopic observations of SN\,2024afav. We provide the light curve morphology and photometric evolution in \S\ref{sec:photometry}, followed by spectroscopic evolution in a \S\ref{sec:specanalysis}. In \S\ref{sec:discussion}, we discuss the unusual features in 2024afav and compare it to objects with similar light curves and spectral properties, and discuss the implications of these features. Finally, we summarize the key results and outline the future direction of this study in \S\ref{sec:conclusion}.

\section{Observations and Data}\label{sec:obs}

\subsection{Discovery}
SN\,2024afav was discovered by the Asteroid Terrestrial-impact Last Alert System~\citep[ATLAS;][]{2018PASP..130f4505T} survey on 2024 December 12.71 UT, with an ATLAS-orange magnitude of 18.3, at R.A.=12$^{\rm h}$49$^{\rm m}$12$^{\rm s}$.5, Decl.= $-$18$^{\circ}$ 06$'$ 12$''$.61 \citep{2024TNSTR5091....1K}. \citet{2025TNSCR.337....1D} classified SN\,2024afav as a hydrogen-poor superluminous supernova at a tentative redshift of $z\approx 0.09$, using spectra obtained on 2025 January 24. We used our high-quality spectra to refine the redshift to $ z = 0.0724\pm 0.0001$ ($d_L\approx 340$ Mpc), which we employ throughout this paper. The SN is located $\approx 0.5''$ from the center of a faint galaxy detected in Legacy survey images, with magnitudes of $m_{g}\approx 22.2$, $m_{r} \approx 21.7$, $m_{i} \approx 21.6$, and $m_{z}\approx 21.6$. At the redshift of the galaxy, the corresponding absolute magnitudes are $M_{g} \approx -15.3$, $M_{r} \approx -15.8$, $M_{i} \approx -16.0$, and $M_{z} \approx -16.0$. Using the SDSS luminosity function from \citet{2009MNRAS.399.1106M}, we infer a host luminosity of $L_g\approx 1.47\times 10^8 L_{g,\odot} \approx 0.011 L^*_{g}$, $L_r\approx 1.50\times 10^8 L_{r,\odot} \approx 0.006L^*_r$.

\subsection{Optical/UV Photometric Observations}

We obtained multi-band photometric observations using the Las Cumbres Observatory network~\citep[LCO;][]{2013PASP..125.1031B}, and the Fred Lawrence Whipple Observatory (FLWO\footnote{\url{https://www.cfa.harvard.edu/facilities-technology/cfa-facilities/fred-lawrence-whipple-observatory-mt-hopkins-az}}) 1.2-m telescope equipped with Keplercam\footnote{\url{https://pweb.cfa.harvard.edu/facilities-technology/telescopes-instruments/12-meter-48-inch-telescope}}. The observations were performed in  $g,r,i,z$ filters. The LCO data were reduced using the \texttt{lcogtsnpipe} pipeline~\citep{2016MNRAS.459.3939V}, while Keplercam data were reduced using a python-based photometric pipeline following standard reduction procedures. In addition to our follow-up observations, we obtained publicly available photometry from the ATLAS. 

We additionally obtained observations with the \swift UV/optical telescope in six filters: UVW2, UVM2, UVW1, $U$, $B$, $V$ (PI: Kumar), and utilized other publicly available UVOT data (PIs: Farah, Schulze).  We performed photometry using \texttt{UVOTSOURCE}\footnote{\url{https://www.swift.ac.uk/analysis/uvot/mag.php}} of \sw{HEASoft}, a part of the Swift software suite, with an aperture of $5''$ radius and standard aperture corrections. 

The entire photometric campaign spans 255 days from the date of discovery (see Table~\ref{tab:photometrytable}). A subset of the optical photometry has been previously presented in \citet{2025arXiv250908051F}. The photometric dataset presented here is in the AB magnitude system for all filters, and has been corrected for Galactic extinction, with $E(B-V) = 0.062$ mag \citep{2011ApJ...737..103S}, assuming the \citet{1999PASP..111...63F} reddening law with $R_V =3.1$. No host extinction correction has been applied. The multi-band light curves are shown in Figure~\ref{fig:lc}. The $c,o$ band and pseudo-bolometric light curves peak at MJD = 60703.3 which we use as Phase = 0 throughout the paper (see Figure~\ref{fig:bolo} and \S\ref{sec:photometry} for details)

\subsection{Optical Spectroscopic Observations}

Following the initial classification of SN\,2024afav as a SLSN-I, we obtained an extensive spectroscopic sequence spanning phases of $+20$ d to +160 d using LCO, the 6.5-m Multi-mirror Telescope (MMT), the 6.5-m Magellan telescopes, and the 8.2-m Gemini-South telescope. Additionally, we included publicly available spectra on WISeREP\footnote{\url{ URL https://wiserep.org }}~\citep{2012PASP..124..668Y} obtained at $-14$ d and $-4$ d phases.

Low-resolution (R $\approx 400$) spectra were obtained with the FLOYDS spectrographs mounted on the 2-m LCO Faulkes telescopes North (FTN) at Haleakalā, and South (FTS) at Siding Spring, Australia. The observations were undertaken as part of the Global Supernova Project from phase +20 d to +91 d. We used a $2\arcsec$ slit along the parallactic angle \citep{Filippenko1982}, covering a wavelength range of $3400 - 10000$ \AA. One-dimensional spectra were extracted, reduced, and calibrated following standard procedures using \texttt{floyds\_pipeline}\footnote{\url{https://github.com/LCOGT/floyds_pipeline}} \citep{Valenti2014}. 

On MMT, we used the Binospec spectrograph \citep{2019PASP..131g5004F} with the LP3800 filter in combination with the 270 lines/mm grating and a $1^{\prime \prime}$-wide slit covering a wavelength range of $3825-9200$ \AA\, with $R\approx 1500$. Two spectra were obtained at phases of +54 d and +139 d. 

We used the Low Dispersion Survey Spectrograph 3~\citep[LDSS3;][]{2016ApJ...817..141S} on the Magellan/Clay 6.5-m telescope with the VPH-All grism and a 1$\arcsec$-wide slit, covering a wavelength range of $4265 - 9650$ \AA\, with $R\approx 700$. Four spectra were obtained at phases spanning +30 d to +160 d. Furthermore, we obtained a high-resolution spectrum using the Magellan Inamori Kyocera Echelle (MIKE), with three exposures of 1800~sec each at $2\times2$ binning and $1^{\prime \prime}$-wide slit resulting in a resolution of $R\approx 19,000$ and $R\approx 25,000$ on the red and blue sides, respectively, covering a wavelength range of $3350 - 9500$ \AA\, at a phase of $+53$ days.

We used the GMOS spectrograph on the Gemini-South telescope to obtain a spectrum at +24 d phase, with the R150 grating, GG455 filter, and a 1$\arcsec$-wide slit, covering a wavelength range of $4744-9999$ \AA\, with $R\approx$300.

The spectra from MMT, Magellan, and Gemini were reduced using the \sw{PypeIt} package~\citep{pypeit:joss_pub} in a standard manner. The one-dimensional spectra were extracted and flux-calibrated using a standard star observation obtained with the same configuration. Data from the high-resolution MIKE spectrograph was reduced using a python-based \sw{CarPy} pipeline~\citep{2000ApJ...531..159K, 2003PASP..115..688K}

\subsection{NIR Spectroscopic Observations}

We obtained NIR spectra with FLAMINGOS2~\citep{2006SPIE.6269E..17E} on Gemini-North and GNIRS~\citep{2006SPIE.6269E..4CE} on Gemini-South. The FLAMINGOS2 spectra were obtained at phases of $-2$ d and $+3$ d using the medium-resolution configuration with a 2-pixel wide slit and an HK filter, covering $1.26 - 2.50$ $\mu$m with $R\approx 900$. The GNIRS spectrum was obtained at a phase of $+23$ d with the short camera, 32 l/mm grating, a 3-pixel-wide slit, and cross-dispersed, covering a wavelength range of $0.82-2.52$ $\mu$m with $R\approx 1200$. The NIR spectra were reduced using the \sw{Pypeit} package, including flat-fielding using GCALflats and sky subtraction obtained on the same night as the spectra. Flux calibration was performed on the extracted spectra using standard star observations obtained on the same night. Finally, a telluric correction was applied to account for atmospheric absorption features.

\begin{figure*}[t!]
\center
\includegraphics[width=0.9\linewidth]{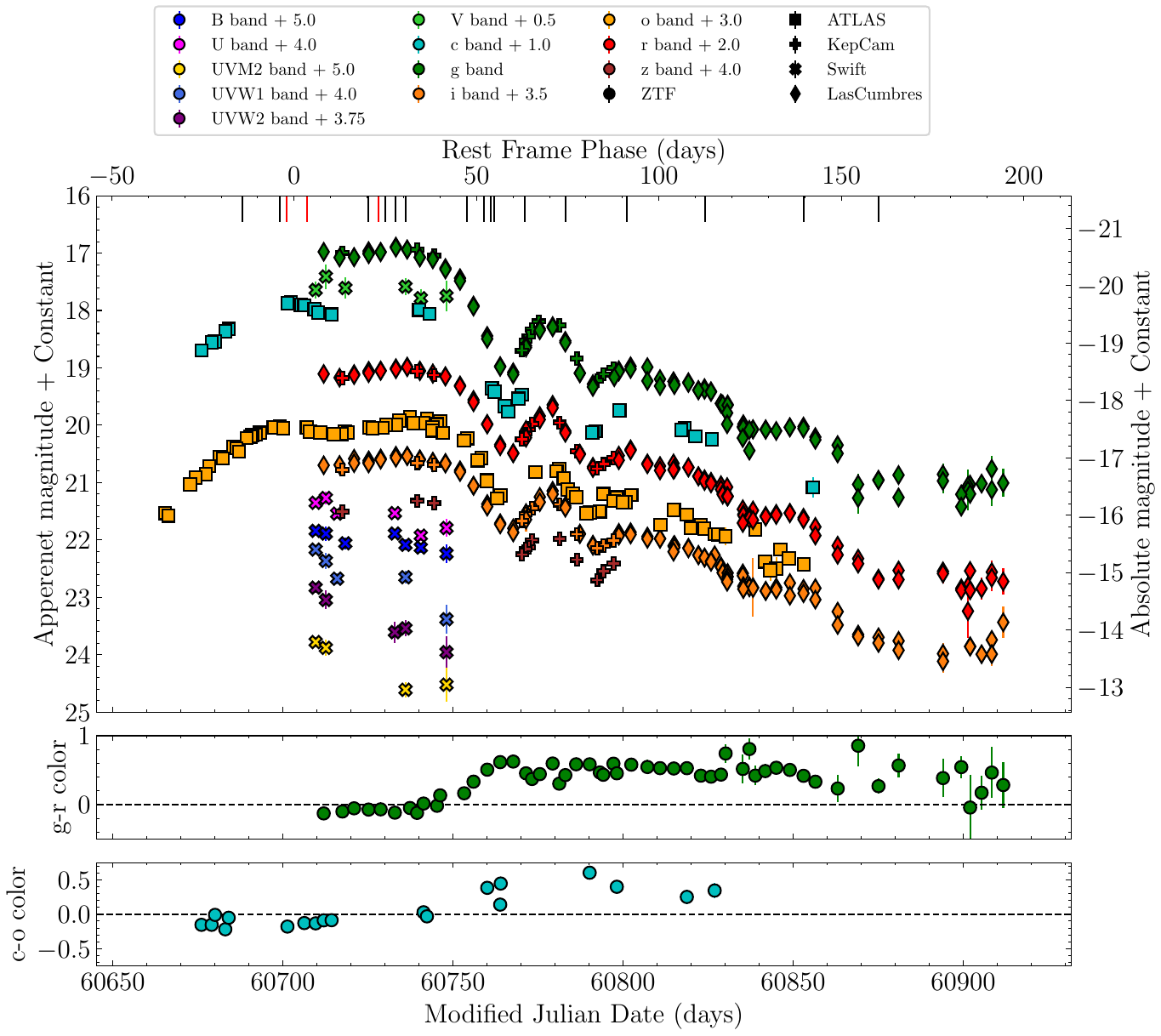}
\caption{Optical/UV light curves of SN\,2024afav with magnitudes in the AB system and corrected for Galactic Extinction. The solid vertical ticks on the top axis mark the epochs of optical (black lines) and near-IR (red lines) spectroscopic observations. The black lines represent optical and the red lines represent epochs of Near-IR spectroscopic observations. The light curve rises from discovery to the first peak in $\approx 46$ d and peaks at an absolute magnitude of $M_{c} \approx -20.7$. Subsequently, it exhibits a second peak of comparable brightness $\approx 38$ d later, before decaying sharply and displaying multiple subsequent undulations over the remainder of its evolution in all bands. The $g-r$ (green) and $c-o$ (blue) colors show an overall evolution from blue to red, but with changes to bluer colors around the light curve peaks.}
\label{fig:lc}
\end{figure*}

\section{Light curve Evolution}
\label{sec:photometry}

\subsection{Light curve Morphology}

The multi-band light curves of SN\,2024afav, shown in Figure~\ref{fig:lc}, exhibit an unusual evolution. The light curve rises for $\approx 46$ days after discovery and peaks at MJD $\approx$ 60703.3 with $M_c\approx -20.70$ and $M_o\approx -20.51$~\citep[K-corrected;][]{2002astro.ph.10394H}. The light curve shows a slight dip in brightness following the initial peak, and then brightens again to a second, equally bright peak at $+40$ d, with $M_c\approx -20.60$, $M_o\approx -20.57$, and $M_g \approx M_r \approx  -20.50$. Following the second peak, the light curve shows a sharp decline, followed by a third peak at $+72$ d with $M_g \approx -19.3$ and $M_r \approx -19.9$. After the first three peaks, the light curve shows four more low-scale undulations at $+94$ with $M_r \approx -19.1$, at $+109$ with $M_r \approx -18.8$, at $+142$ with $M_r \approx -18.0$, and at $+175$ with $M_r \approx -17.0$. The light curve starts to show the start of another peak; however, no photometric coverage is available due to visibility constraints. The peak absolute magnitude is at the 78th percentile of the SLSNe-I population~\citep{2024MNRAS.535..471G}.

The $g-r$ and $c-o$ colors are initially $\approx -0.2$ to 0 mag during the first two peaks, and then redden rapidly after the second peak to a mean value of $\approx +0.6$ mag. However, the densely-sampled $g-r$ color curve exhibits changes to a bluer color at the time of subsequent light curve peaks.

\begin{longtable*}{|c|c|c|c|c|}
\hline
\hline
Peak\,Number & MJD & Phase & $L_{\text{bol}}$ & FWHM \\
             &     & (days) &  (erg s$^{-1}$) & (days) \\
\hline
1   &  60703.3 & 0 &  $7.48 \times 10^{43}$   & 23.0    \\
2  &  60745.7  & +40 & $5.41 \times 10^{43}$ &  29.6  \\
3 &  60781.1  & +72 & $2.24 \times 10^{43}$  & 8.8   \\
4 & 60803.7  &  +94 & $1.25 \times 10^{43}$ &  12.1   \\
5 & 60819.8  & +109 & $0.96 \times 10^{43}$ & 11.6  \\
6 &  60851.9 & +142 & $0.48 \times 10^{43}$ & 12.5   \\
7 & 60890.5 & +175 & $0.18 \times 10^{43}$ & 14.1 \\
\hline
\caption{The properties of significant peaks in the pseudo-bolometric light curve.} \label{tab:peaklist}  \\
\end{longtable*}

\subsection{Temperature and Radius Evolution} 
\label{sec:bolo}

\begin{figure}[t!]
\center
\includegraphics[width=0.6\linewidth]{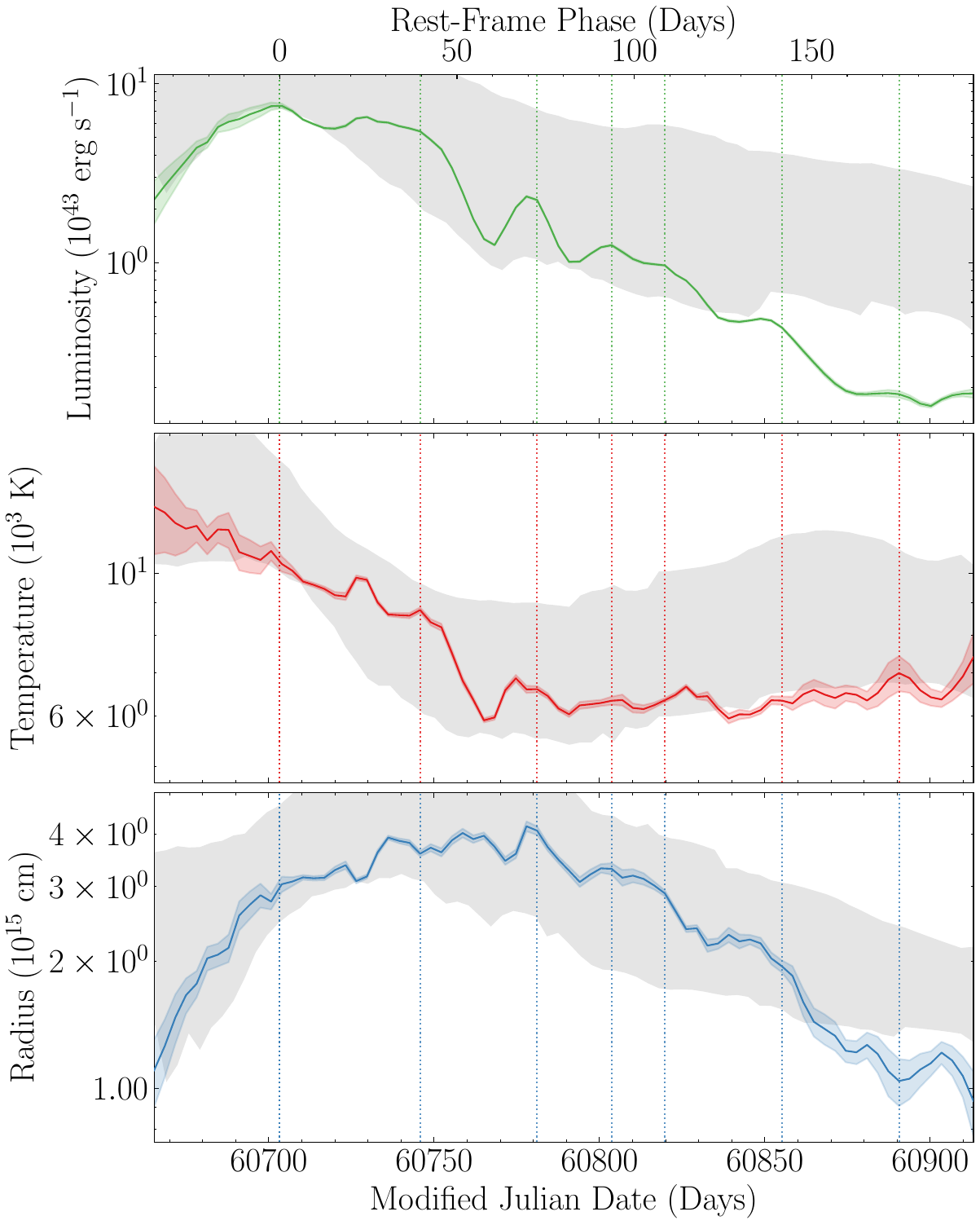}
\caption{The pseudo-bolometric light curve (top), photospheric temperature (middle), and photospheric radius (bottom) of SN\,2024afav. The grey-shaded regions represent the range of parameters for quantities in the SLSN-I population, as reported by \citet{2024MNRAS.535..471G}. The pseudo-bolometric light curve exhibits multiple bumps (marked by vertical lines). The peak luminosity and width of each peak are provided in Table~\ref{tab:peaklist}. The temperature and radius fluctuations closely follow the peaks.}
\label{fig:bolo}
\end{figure}

We used the \sw{extrabol} \sw{python} package~\citep{2024RNAAS...8...48T} to determine the bolometric light curve, photospheric temperature, and radius evolution of SN\,2024afav, assuming a blackbody spectrum. The results are shown in Figure~\ref{fig:bolo}. We fit a smooth curve to the overall light curve using a large bin size of 30 to minimize the effects of undulation on the fitted curve. This trend was then subtracted from the bolometric light curve and \sw{scipy.signal.find\_peaks} \sw{python} package was used to get the peaks in the light curve. Including the primary peak, the light curve shows a total of seven peaks in the bolometric light curve (see Table~\ref{tab:peaklist}). We estimate a peak luminosities of $\approx 7.5 \times 10^{43}$ erg s$^{-1}$ for primary peak. The positions, L${_\text{p, bol}}$ and half widths, for all peaks have been listed in Table~\ref{tab:peaklist}. Integrating the light curve, we find a total radiated energy of $\approx 5.1 \times 10^{50}$~erg, on the lower end of the SLSNe-I population~\citep{2024MNRAS.535..471G}.

The temperature declines from $\approx 12,700$ K at the first epoch to $\approx 10,340$ K until the primary peak. Thereafter, the temperature decreases further to $\approx 8,500$ K around the secondary peak, before showing a sudden decline to $\approx 5,900$ K at $+60$ d phase. After this epoch, the temperature shows a rise to $\approx 6,870$ K, following the third peak in the light curve at $\approx +70$ d phase. After the third peak, the temperature evolution shows small fluctuations and closely follows the light curve undulations and remains close to $\approx 6,000$ K.

The photospheric radius shows an increase from $\approx 10^{15}$ cm at first detection to $\approx 3 \times 10^{15}$ cm at the primary peak. Subsequently, the radius continues to rise more gradually past the secondary peak to a maximum value of $\approx 3.9 \times 10^{15}$ cm at $+40$ d. The radius then declines back to $\approx 10^{15}$ cm at $+180$ d.  Throughout this overall evolution, the photospheric radius exhibits smaller fluctuations during the multiple light curve peaks.

\begin{figure}[t!]
\centering
\subfigure{
    \includegraphics[width=0.45\linewidth]{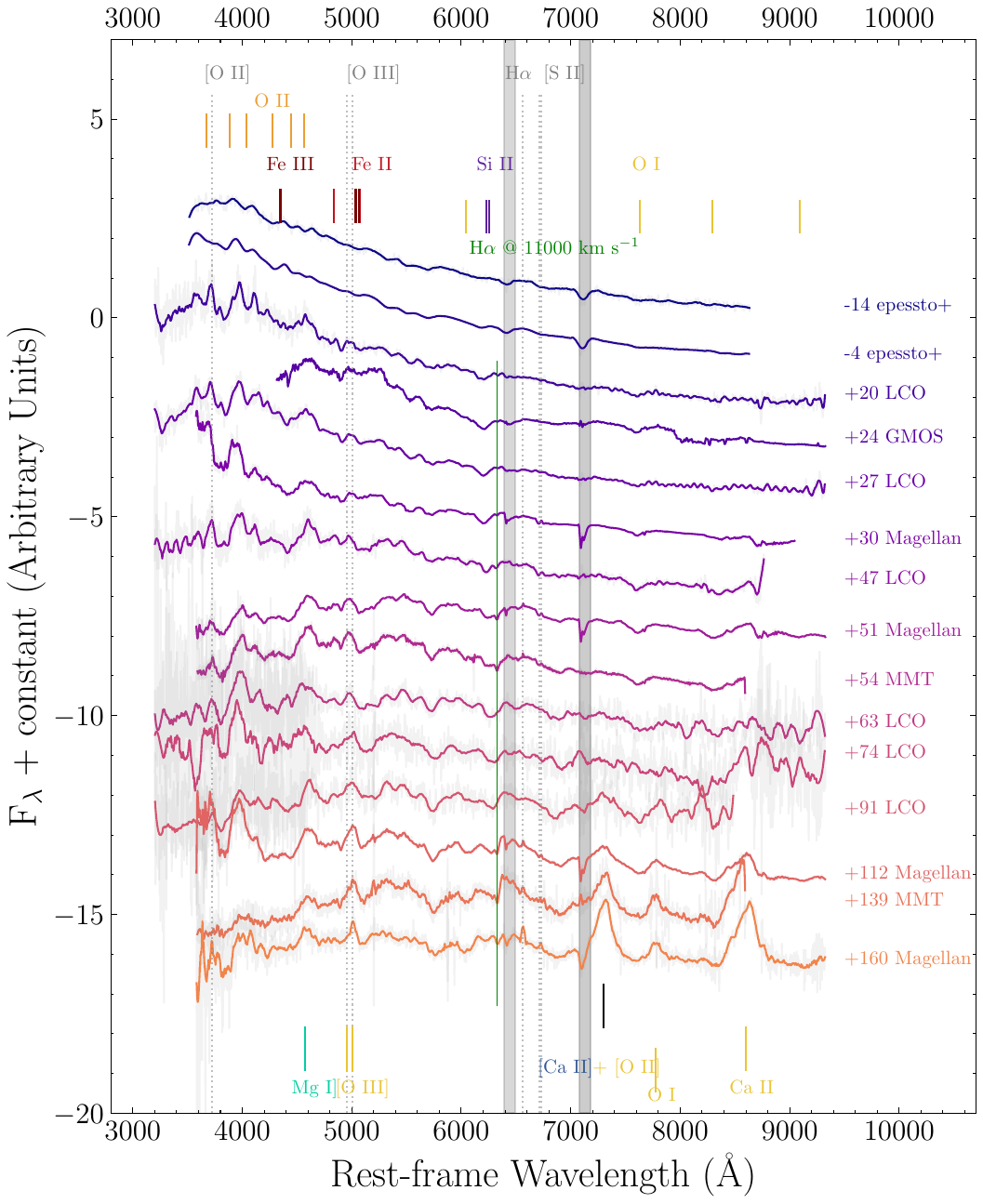}
}
\subfigure{
\includegraphics[width=0.45\textwidth]{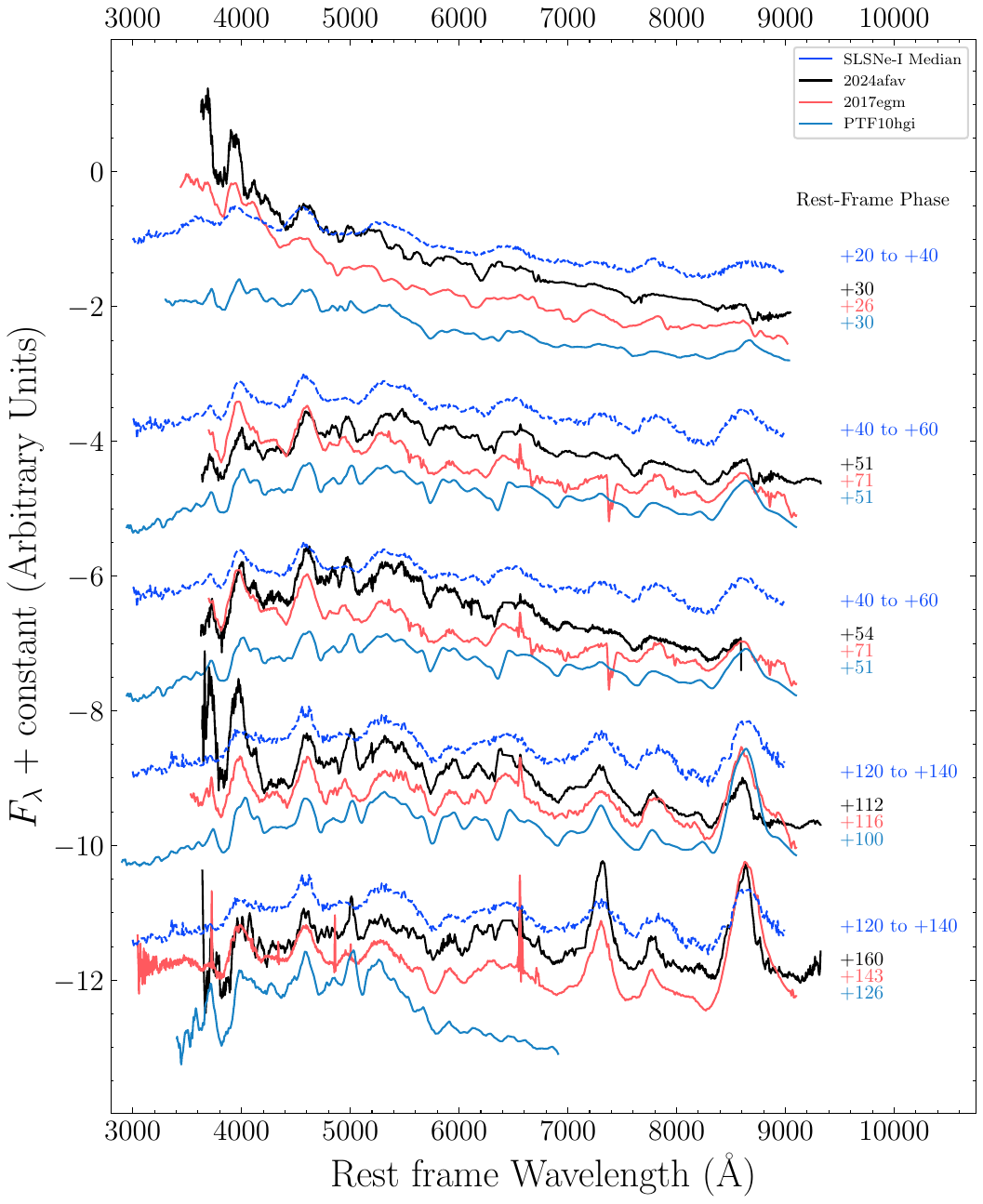}
}
\caption{{\it Left:} Optical spectra of SN\,2024afav covering phases of $-14$ d to +160 d. The early-time spectra are blue, with characteristic \ion{O}{2} features. In the post-peak phases we observe other common SLSNe-I features of \ion{Fe}{2}, \ion{Fe}{3}, \ion{Na}{1}D, \ion{Si}{2} and \ion{O}{1}, along with unusual narrow features such as [\ion{O}{3}] emission and H$\alpha$ absorption. The $\gtrsim +112$ d spectra show nebular NIR \ion{Ca}{2} triplet and [\ion{Ca}{2}] + [\ion{O}{2}] complex. The colored spectra are smoothed using the Savitzky–Golay filter~\citep{2018JPhCS1141a2151S}. {\it Right:} Comparison of SN\,2024afav spectra to SN\,2017egm, PTF10hgi, and median SLSNe-I spectra \citep{2025arXiv250321874A} at several phases. SN\,2024afav shows remarkable similarity to SN\,2017egm, PTF10hgi and SN\,2019hge including the unusual [\ion{O}{3}] and [\ion{Ca}{2}] + [\ion{O}{2}] complex emission features. The narrow H$\alpha$ feature in the SN\,2024afav spectra is at a similar velocity and width to H$\alpha$ in PTF10hgi \citep{2020ApJ...902L...8Y}. A detailed identification of all features in the post-peak phase is provided in Figure~\ref{fig:specident}.}
\label{fig:spec}
\end{figure}

\begin{figure}[t]
\center
\includegraphics[width=0.9\linewidth]{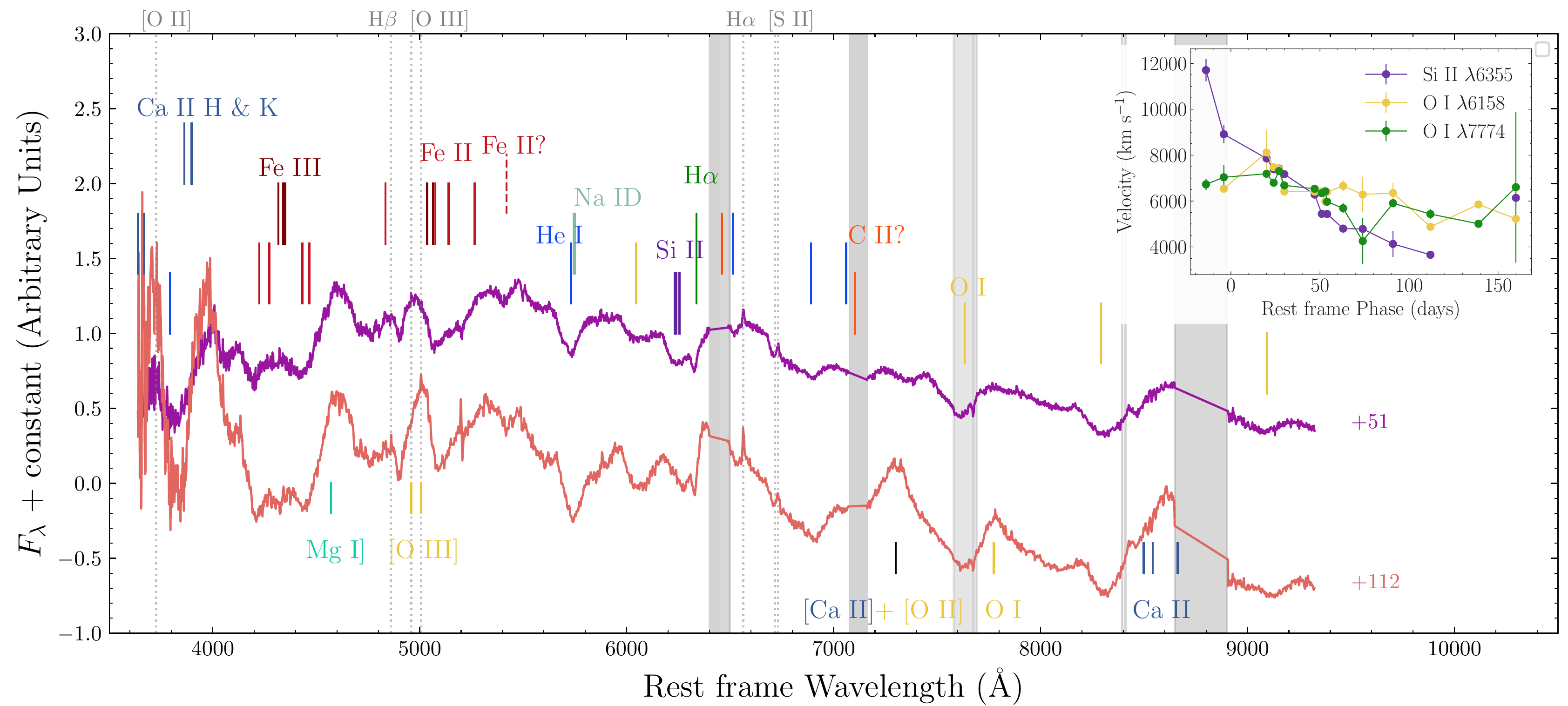}
\caption{Detailed identification of spectral features in the optical spectra at $+51$ and $+112$ d. The majority of the absorption features are marked above the spectra using a blueshifted velocity of 7,500 km s$^{-1}$, while the \ion{Fe}{2}, \ion{Fe}{3} and \ion{Si}{2} are at 5,500 km s$^{-1}$. The SN emission features are marked below the spectra at their rest frame wavelengths.  The host galaxy emission lines are marked with vertical gray dotted lines.  The inset shows the photospheric velocity evolution estimated using relatively isolated features of \ion{Si}{2} $\lambda 6355$ \AA, \ion{O}{1} $\lambda 6158$ \AA, and \ion{O}{1} $\lambda 7774$ \AA. 
} 
\label{fig:specident}
\end{figure}

\section{Spectroscopic Analysis} 
\label{sec:specanalysis}

\subsection{Optical Spectral Evolution Overview}

The complete optical spectral sequence from phase $-14$ to $+160$ d is shown in Figure~\ref{fig:spec}, and a detailed identification of features using SN\,2017egm, PTF10hgi and median spectra from ~\citet{2025arXiv250321874A}, is marked in Figure~\ref{fig:specident} at the $+51$ d and $+112$ d phases. The spectral evolution follows a typical trend observed in SLSNe-I, from a hot, blue continuum with absorption lines at early times to a cooler, emission-line-dominated spectrum in the near-nebular phase. The early pre-peak spectra exhibit the typical ``w''-shape broad features of \ion{O}{2} in the $3500 - 5000$\AA\, region with a velocity $\approx$11,600 km s$^{-1}$. The early spectra also show features in the $5000-6000$ \AA\, region commonly associated with \ion{Na}{1}D and \ion{Si}{2} and sometimes \ion{He}{1} \citep{2023ApJ...943...41C, 2024MNRAS.535..471G, 2025MNRAS.541.2674A}.

Post-peak, as the temperature declines to $\lesssim 10^4$ K, the \ion{O}{2} lines weaken and the $3500 - 5000$ \AA\ region is dominated by \ion{Fe}{2} and \ion{Fe}{3} lines with noticeable presence of $\lambda$5169 \AA\, feature. The $5000-6500$ \AA\, region shows three absorption features that increase in strength over time, which we identify as \ion{He}{1} $\lambda 5876$ \AA\, and \ion{Na}{1}D $\lambda \lambda 5890, 5896$ \AA\, complex, \ion{O}{1} $\lambda \lambda 6150$ \AA\, and \ion{Si}{2}$\lambda 6355$ \AA. Meanwhile, \ion{O}{1} $\lambda 7774$ \AA\, appears in the $+20$ d spectrum and becomes stronger over time until $+63$ d, where it starts to show an associated emission component. From $+20$ d to $+74$ d, the velocity decreases from $\approx 7,500$ km s$^{-1}$ to $\approx 6,300$ km s$^{-1}$, estimated using isolated \ion{Si}{2} and \ion{O}{1} features.

Starting at a phase of $+91$ d, the spectra begin to exhibit near-nebular emission lines of \ion{Mg}{1}] $\lambda$ 4571 \AA\, and the \ion{Ca}{2} $\lambda \lambda \lambda 8498, 8542, 8662$ \AA\, NIR triplet. Furthermore, the \ion{O}{1} $\lambda 7774$ \AA\, feature changes from absorption to emission and becomes stronger over time until our last spectrum at $+160$ days. We estimate a velocity of $\approx 6,000$ km s$^{-1}$ at a phase of $\approx 100$ d using \ion{O}{1} $\lambda$ 6158 \AA\, and $\lambda$ 7774 \AA. Moreover, we find an emission feature of [\ion{Ca}{2}] and [\ion{O}{2}] centered at $\approx 7300$ \AA\, starting at $+112$ d, which becomes stronger thereafter. This feature is discussed in detail in \S\ref{sec:novelfeat}.

SN\,2024afav spectral features exhibit lower velocities than the median spectra of SLSNe-I \citep{2025arXiv250321874A}; as shown in Figure~\ref{fig:spec}. The spectral evolution, features, and velocity progression most closely resemble those of the unusual events SN\,2017egm and PTF10hgi \citep{2023ApJ...949...23Z, 2020ApJ...902L...8Y}, which share a few distinctive features in their optical spectra. We discuss the distinctive feature in detail in \S~\ref{sec:novelfeat} and \S~\ref{sec:discussion}. The spectral similarities from peak to post-peak phases make SN\,2024afav a spectral analog of PTF10hgi and SN\,2017egm events.


\subsection{Unusual Spectral Features}
\label{sec:novelfeat}

\begin{figure}[t!]
\centering
\subfigure{
    \includegraphics[width=0.8\linewidth]{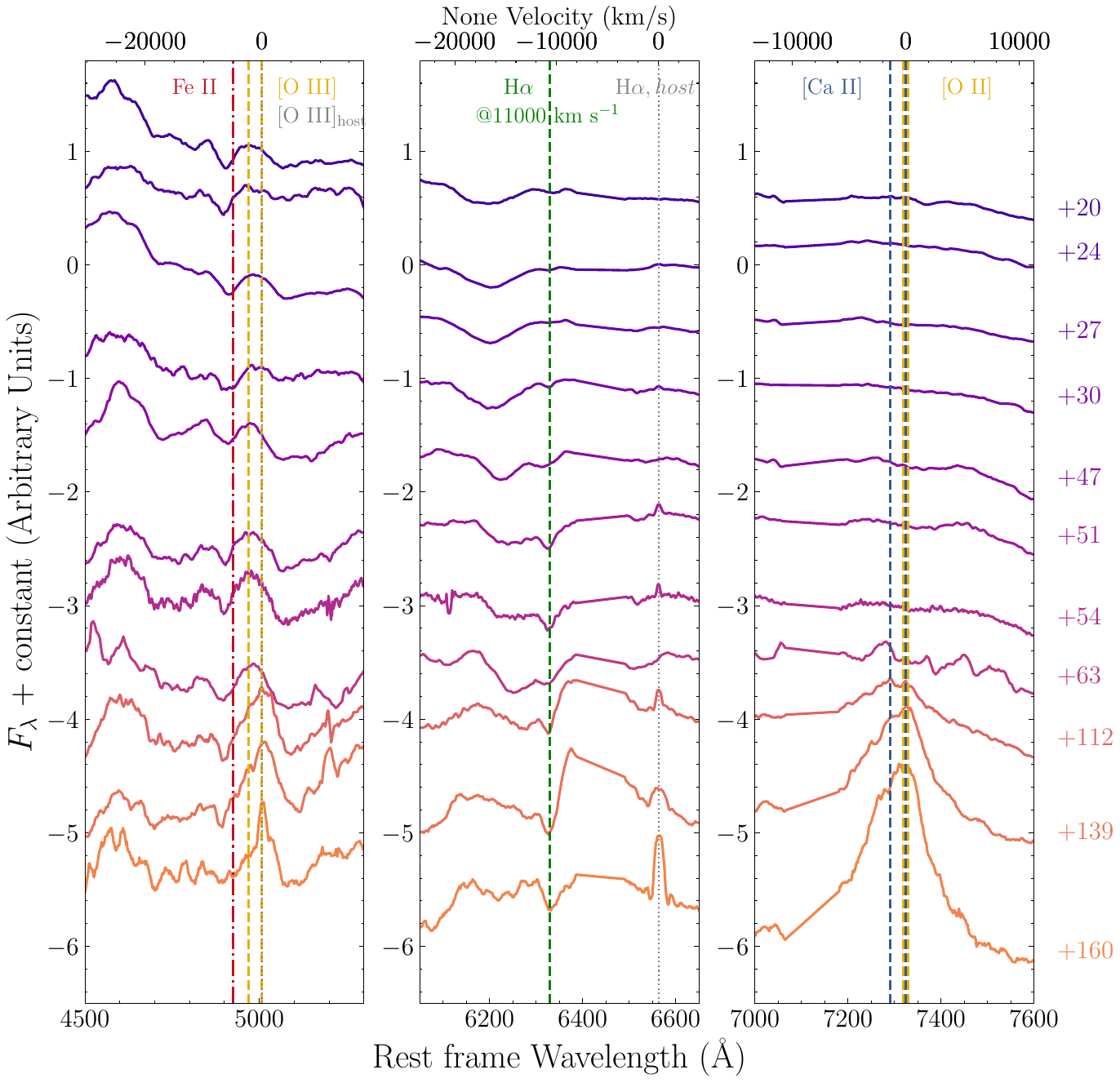}
}
\subfigure{
    \includegraphics[width=0.9\linewidth]{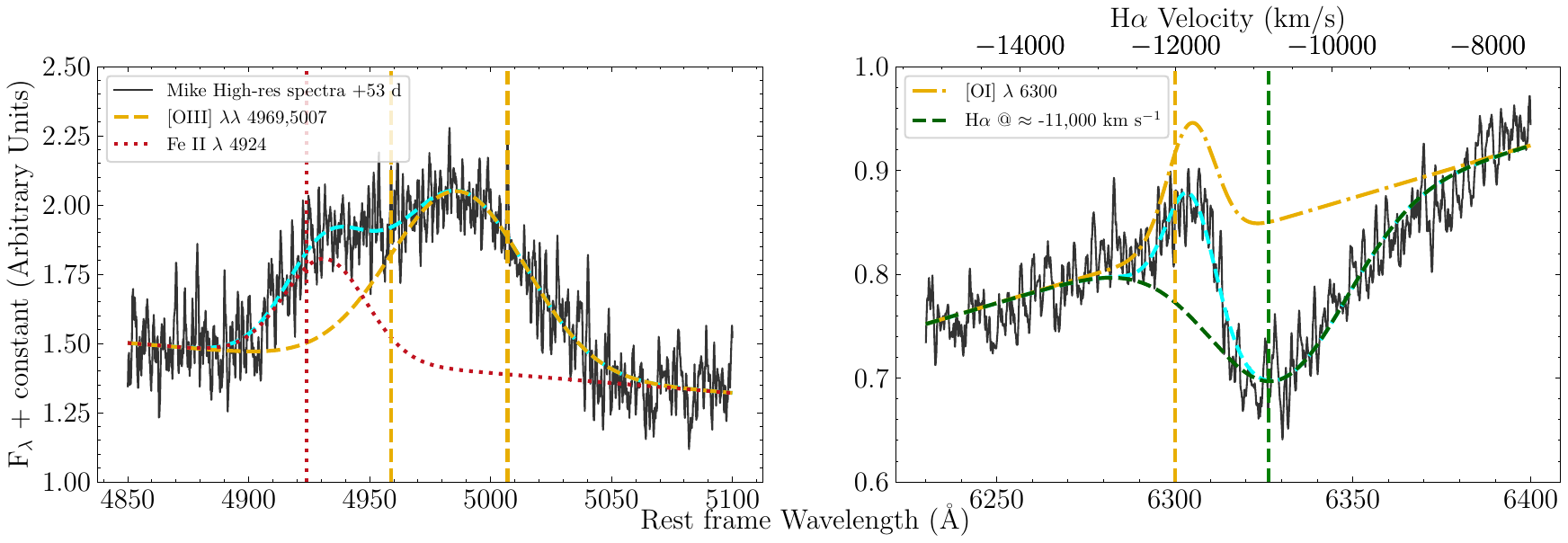}
}
\caption{{Top:} Evolution of the unusual spectral features [\ion{O}{3}] ({\it left}), H$\alpha$ ({\it middle}) and the [\ion{O}{2}] + [\ion{Ca}{2}] complex ({\it right}). The [\ion{O}{3}] feature exhibits contamination from \ion{Fe}{2} in early spectra and displays a shift towards [\ion{O}{3}] over time. The H$\alpha$ feature (green dashed line) is narrow ($1,800$ km s$^{-1}$) and remains at a constant blueshift velocity of $\approx 11,000$ km s$^{-1}$. The [\ion{O}{2}] + [\ion{Ca}{2}] complex appears at $\approx +112$ d and becomes stronger over time. This feature is centered on [\ion{O}{2}] implying the stronger presence of [\ion{O}{2}] compared to [\ion{Ca}{2}].  Telluric absorption is masked in the middle and right panels. {\it Bottom:} Line fits to the [\ion{O}{3}] $\lambda \lambda 4969, 5007$ \AA\, ({\it left}) and H$\alpha$ ({\it right}) profiles in the MIKE echelle spectrum at $+53$ d. The fitted profile shows a clear presence of [\ion{O}{3}] and a contribution from \ion{Fe}{2} $\lambda 4924$ \AA. The H$\alpha$ profile exhibits asymmetric absorption, which is due to emission from [\ion{O}{1}] $\lambda 6300$ \AA.}
\label{fig:novelfeat}
\end{figure}

While most of the spectral features in SN\,2024afav are typical of SLSNe-I, some are unusual; we show the evolution of these features in Figure~\ref{fig:novelfeat} and discuss them in detail below. 

\subsubsection{[\ion{O}{3}] $\lambda \lambda 4969, 5007$ \AA}

From $+20$ d onwards, we observe an unusual emission feature at $\approx 4940$~\AA\, with a possible absorption component centered at $\approx 4880$ \AA. The overall profile of the emission feature changes with time, with the peak shifting to a wavelength of $\approx 5000$~\AA\, at later epochs (left panel of Figure~\ref{fig:novelfeat}). While it is plausible to associate this unusual absorption and emission feature with \ion{Fe}{2} $\lambda$4924 \AA, a detailed study of the SLSN-I sample shows that \ion{Fe}{2} $\lambda$5169 \AA\, is stronger than \ion{Fe}{2} $\lambda$4924 \AA\, at all phases \citep{2025arXiv250321874A}, but We do not find a strong \ion{Fe}{2} $\lambda$5169 \AA\, feature in SN\,2024afav. This, and the shift in line centroid, indicate that this feature is not predominantly due to \ion{Fe}{2} $\lambda$4924 \AA, and instead has a significant contribution from the [\ion{O}{3}] $\lambda \lambda 4969, 5007$ \AA\, doublet. Using our high-resolution MIKE spectrum at $+53$ d we fit the emission profile and find the definitive presence of [\ion{O}{3}] $\lambda \lambda 4969, 5007$ \AA, as well as a contribution from \ion{Fe}{2} $\lambda$4924 \AA\, (Figure~\ref{fig:novelfeat}). 
Based on the overall evolution from early to later phase, we conclude that at $\lesssim +50$ d this feature is a complex of \ion{Fe}{2} $\lambda$ 4929 \AA\, and [\ion{O}{3}] $\lambda \lambda 4959,5007$ \AA. At $\gtrsim +50$ d, the [\ion{O}{3}] feature becomes stronger, explaining the wavelength shift and enhanced strength of this emission feature compared to typical SLSNe-I.

\subsubsection{H$\alpha$}
We observe a narrow absorption feature centered at $\approx 6330$ \AA\, starting at $+30$ d (and potentially even at $+20$ d; center panel of Figure~\ref{fig:novelfeat}). We identify this feature as a potential high velocity H$\alpha$ absorption using a spectral match to PTF10hgi \citep{2020ApJ...902L...8Y}. This feature retains a nearly constant velocity of $\approx -11,000$ km s$^{-1}$ and narrow FWHM width of $\approx 1,800$ km s$^{-1}$. Using our high-resolution MIKE spectrum at $+53$ d, we fit the profile of this feature and find that it requires an additional weak emission component on the bluer side (bottom-right panel of Figure~\ref{fig:novelfeat}), which we identify as a potential \ion{O}{1} $\lambda 6300$ \AA\, feature. We conclude that the absorption feature is a narrow, blueshifted H$\alpha$, retaining constant velocity and width throughout the spectral sequence.


\subsubsection{[\ion{O}{2}] $\lambda \lambda 7319, 7330$  \AA\, and [\ion{Ca}{2}] $\lambda \lambda 7292, 7324$ \AA}

A distinctive emission profile emerges at $\approx 7300$ \AA\, at $+112$ d (right panel of Figure~\ref{fig:novelfeat}). This feature shows some asymmetry on the bluer end due to the presence of \ion{He}{1} $\lambda 7281$ \AA. We fit a multi-Gaussian profile and find three components in the line profile: a narrow component centered at 7330 \AA\, a wider component at 7300 \AA, and a broad, lower-intensity component fitting the continuum. We identify these features as the [\ion{O}{2}] $\lambda \lambda 7319, 7330$  \AA\, and [\ion{Ca}{2}] $\lambda \lambda 7292,7324$ \AA\, doublets. The center of the overall profile is closer to the [\ion{O}{2}] doublet, implying that the feature is dominated by [\ion{O}{2}]. Additionally, this feature becomes stronger at the same time when the [\ion{O}{3}] $\lambda 5007$ \AA\, feature becomes weaker at $\gtrsim + 100$ d, implying a transition from [\ion{O}{3}] to [\ion{O}{2}]. For our spectra at $\gtrsim +112$ d, we estimate the L$_{7300}$/L$_{6300}$ ratio $\approx 4.82\pm 1.14$ using the $7300$ \AA\, and $6300$ \AA\, features. The L$_{7300}$/L$_{6300}$ ratio for SN\,2024afav is greater than the typical values estimated from other SLSNe-I~\citep{2019ApJ...871..102N, Blanchard2025}. Furthermore, the ratio estimated from the $\gtrsim + 100$ d phase spectra of SN\,2024afav exhibits an increasing trend (see Table~\ref{tab:featstrength}), similar to some SLSNe-I such as SN\,2017egm, PTF10hgi, and LSQ14an that show possible CSM interaction~\citep{2019ApJ...871..102N,Blanchard2025}.

\subsection{NIR spectroscopic analysis} \label{sec:NIRanalysis}

\begin{figure}[t!]
\centering
\includegraphics[width=0.95\linewidth]{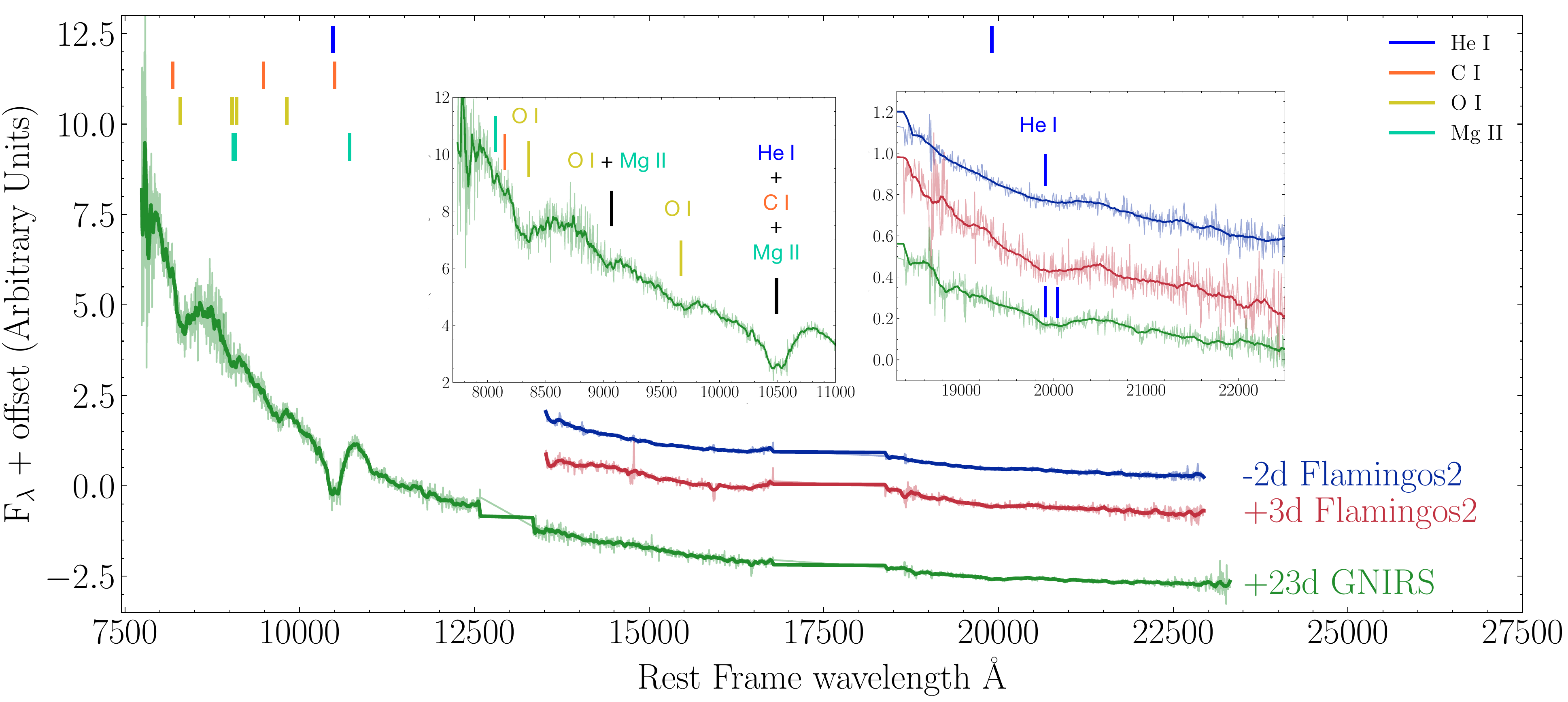}
\caption{Near-IR spectra at $-2$, $+3$, and $+23$ d. The insets show zoomed-in versions of the spectra in the 1 and 2 $\mu$m regions. All three spectra show \ion{He}{1} $\lambda$2.058 $\mu$m, implying the presence of helium in the ejecta. The 1 $\mu$m region exhibits broad features of \ion{O}{1}, \ion{C}{1}, and \ion{Mg}{2} in addition to \ion{He}{1} $\lambda$1.083 $\mu$m. We find that both \ion{He}{1} $\lambda$1.083 $\mu$m and $\lambda$2.058 $\mu$m at $+23$ d exhibit a ``w''-shaped profile, potentially indicative of CSM interaction.
The blueshift velocity of \ion{He}{1} $\lambda$2.058 $\mu$m feature is 10,500 km s$^{-1}$ at $-2$ d, 10,000 km s$^{-1}$ at $+3$ d, and 10,500 km s$^{-1}$ and 7,500 km s$^{-1}$ for the two minima at $+23$ d for faster and slower components respectively. In $+23$ d spectra \ion{He}{1} $\lambda$1.083 $\mu$m feature has same velocity as $\lambda$2.058 $\mu$m features for both components.}
\label{fig:NIRspec}
\end{figure}

We show the NIR spectra of SN\,2024afav in Figure~\ref{fig:NIRspec}. The near-peak NIR spectra obtained at $-2$ d and $+3$ d cover $\approx 1.35-2.3$ $\mu$m and exhibit a single broad absorption feature, corresponding to \ion{He}{1} $\lambda2.058$ $\mu$m (\ion{He}{1} 2-$\mu$m line hereafter) with a blueshifted velocity of $-10,500$ km s$^{-1}$. No other features are detected, for example, \ion{S}{1}, \ion{Mg}{1}, and \ion{Si}{1} which are commonly observed in Type Ib/c \citep{2022ApJ...925..175S}. 



The next NIR spectrum at $+23$ d, covering $0.78-2.3$ $\mu$m 
exhibits the same \ion{He}{1} 2-$\mu$m absorption, but with an apparently bimodal velocity structure, with velocities of $7,500$ and $10,500$ km s$^{-1}$ (inset of Figure~\ref{fig:NIRspec}). 
The strongest feature in the spectrum, at $\approx 1.05$ $\mu$m is a blend of \ion{Mg}{2} $\lambda$1.095 $\mu$m, \ion{He}{1} $\lambda$1.083 $\mu$m, and possibly \ion{C}{1} $\lambda$1.069 $\mu$m  (see Figure~\ref{fig:NIRspec}). This absorption feature also shows two distinctive absorption troughs at $\approx 1.045$ $\mu$m and $\approx 1.055$ $\mu$m, mirroring the profile of the \ion{He}{1} 2-$\mu$m line. The possible blend of several features in this wavelength range makes the interpretation of the line profile complicated, but the similarity to the unblended \ion{He}{1} 2-$\mu$m line suggests that it is dominated by \ion{He}{1} $\lambda$1.083 $\mu$m at two different velocities.


Additionally, it shows a series of absorption features centered at $\approx 8250, 9000, 9700$ \AA. These features correspond to \ion{O}{1} $\lambda 8446$ \AA, $\lambda 9263$ \AA\, with some possible contribution from \ion{C}{1} $\lambda 8335$ \AA, $\lambda 9405$  \AA\, and \ion{Mg}{2} $\lambda \lambda 9218$, 9244 \AA\, features; however, since we do not see conclusive evidences of carbon lines in the optical spectra and no signs of \ion{Mg}{2} $\lambda 21062$ \AA\, features, suggesting the NIR features are likely dominated by oxygen. Regardless of the interpretation, we conclusively detect \ion{He}{1}, providing strong evidence of its presence in ejecta of SN\,2024afav.

\section{Discussion}  
\label{sec:discussion}

\subsection{Spectroscopic Evidence for Post-Peak CSM Interaction}

The presence of narrow and blueshifted H$\alpha$ absorption, persistent \ion{He}{1} optical and NIR lines, and early appearance of forbidden high-ionization oxygen line emission ([\ion{O}{2}] and [\ion{O}{3}]) are highly unusual for SLSNe-I. By definition, these events do not show signatures of hydrogen, with the exception of a few events that, unlike SN\,2024afav, showed hydrogen only at very late time \citep{2017ApJ...848....6Y}. The \ion{He}{1} features have similarly only been detected in a few SLSNe-I to date, generally near peak when a small amount of helium retained by the progenitor is non-thermally excited by the central magnetar \citep{2016MNRAS.458.3455M,2019ApJ...871..102N,2024MNRAS.535..471G}. These features weaken in the post-peak phase when the ejecta expand and the engine energy input declines, as seen recently in SN\,2024rmj \citep{2025ApJ...987..127K}.  The late-time presence of \ion{He}{1} in SN\,2024afav requires a different excitation mechanism, likely CSM interaction (e.g., \citealt{2020ApJ...902L...8Y, 2024A&A...692A.204D}). Similarly, when detected, the [\ion{O}{2}] and [\ion{O}{3}] lines are generally observed only in the late nebular phase, indicative of ionization of the inner ejecta by the central engine \citep{2019ApJ...871..102N, Blanchard2025}. In a few cases, the early appearance of [\ion{O}{3}] has been interpreted as a signature of CSM interaction  \citep{2024MNRAS.52711970A,2024A&A...683A.223S}. We thus conclude that the appearance and properties of these features are indicative of CSM interaction in the post-peak phase.

The narrow width, large blueshift, and appearance starting only at +20 d of H$\alpha$ absorption are distinct from other spectral features in SN\,2024afav (Figure~\ref{fig:specident}), suggesting that hydrogen is not part of the initial SN ejecta and was instead swept up by the fast ejecta from the surrounding CSM. The lack of an H$\alpha$ emission component further indicates that the hydrogen was most likely present in a narrow shell rather than in a diffuse CSM, commonly seen in interacting SNe such as SN IIn \citep{2016MNRAS.458.2094D, 2017hsn..book..875C, 2022A&A...660L...9D, 2024arXiv240504259D}. 

The broad optical and NIR helium in the pre- and near-peak spectra (before the appearance of narrow H$\alpha$ absorption) indicate that the bulk of the helium is present in the ejecta, and therefore that the progenitor star retained a helium layer prior to explosion. On the other hand, the clear post-peak bimodal velocity structure of \ion{He}{1} $\lambda2.058$ $\mu$m, with an emerging component at the same velocity as H$\alpha$ and at the same phase, indicates that helium is also present in the CSM.  The optical helium features persist until our last spectrum at +160 d, indicating that the CSM interaction-based excitation continues to this late phase.

The [\ion{O}{2}] $\lambda \lambda 7320,7330$ \AA feature is rare and observed in some of SLSNe-I only in the nebular phase when the ejecta is nearly transparent, revealing the inner regions \citep{2019ApJ...871..102N}. Moreover, the [\ion{O}{3}] $\lambda \lambda 4959, 5007$ \AA\ feature is seldom seen and requires an excitation mechanism to produce a higher ionization state and a low-density region in ejecta. In SN\,2024afav, the early appearance of [\ion{O}{3}] starting at about +50 d and [\ion{O}{2}] at +110 d, requires such a region in the outer part of ejecta, as the rest of the spectrum suggests that the ejecta has not reached a nebular state. Such a low-density and high-ionization region in the outer ejecta can be produced by the interaction of ejecta with the CSM, explaining these features at such early epochs \citep{2024MNRAS.52711970A}.

Finally, we note that in addition to the emergence of the various interaction-dominated spectral features at about +20 d, the light curve begins to exhibit its first bump, and the photospheric temperature exhibits an increase, at the same phase -- indicators of the same CSM interaction. 

Combining the available spectroscopic and photometric information, we can provide a rough picture of the CSM. The multiple bumps (and lack of H$\alpha$ emission) point to a patchy or shell-like CSM rather than a smooth wind-like CSM.  The CSM is hydrogen- and helium-rich. Using the fastest velocity of $\approx 12,000$ km s$^{-1}$ for H$\alpha$ at $+20$ d, and an explosion date at $\approx -49$ d \citep{2025arXiv250908051F}, we estimate that the nearest CSM shell is located at $\approx 7 \times 10^{15}$ cm; for a CSM velocity of $\sim 1,000$ km s$^{-1}$, this shell would have been ejected $\approx 2.2$ yr before explosion.  The CSM shell is swept by the ejecta, producing a forward shock \citep{2017hsn..book..403S} that accelerates the CSM to the ejecta velocity, thereby creating the blueshifted narrow H$\alpha$ absorption (and high-velocity component of \ion{He}{1}) as the shocked region is viewed above the photosphere. This layer encounters no obstacles in the absence of a diffuse CSM until another shell is encountered, at which point it maintains the same velocity (and produces additional bumps in the light curve). The reverse shock produced during the CSM interaction ionizes the outer layers of the ejecta, which are primarily composed of oxygen and helium.  This helps to sustain the excitation of helium, and moreover, low-density regions in these layers produce the [\ion{O}{3}] emission. Subsequently, as the ejecta expand, the low-density regions transition to a lower ionization state, explaining the emergence of [\ion{O}{2}] at a later phase than [\ion{O}{3}]. 

To summarize, post-peak CSM interaction provides a consistent explanation for the unusual spectral features, light curve bumps, and their temporal correlation. The interaction-induced forward shock explains the presence of a narrow H$\alpha$ (and \ion{He}{1}), while the reverse shock provides a source of helium excitation and oxygen ionization in the outer ejecta. 

\subsection{Comparison to Previous Multi-Peaked SLSNe-I}

\begin{figure}[t!]
\centering
\includegraphics[width=0.65\linewidth]{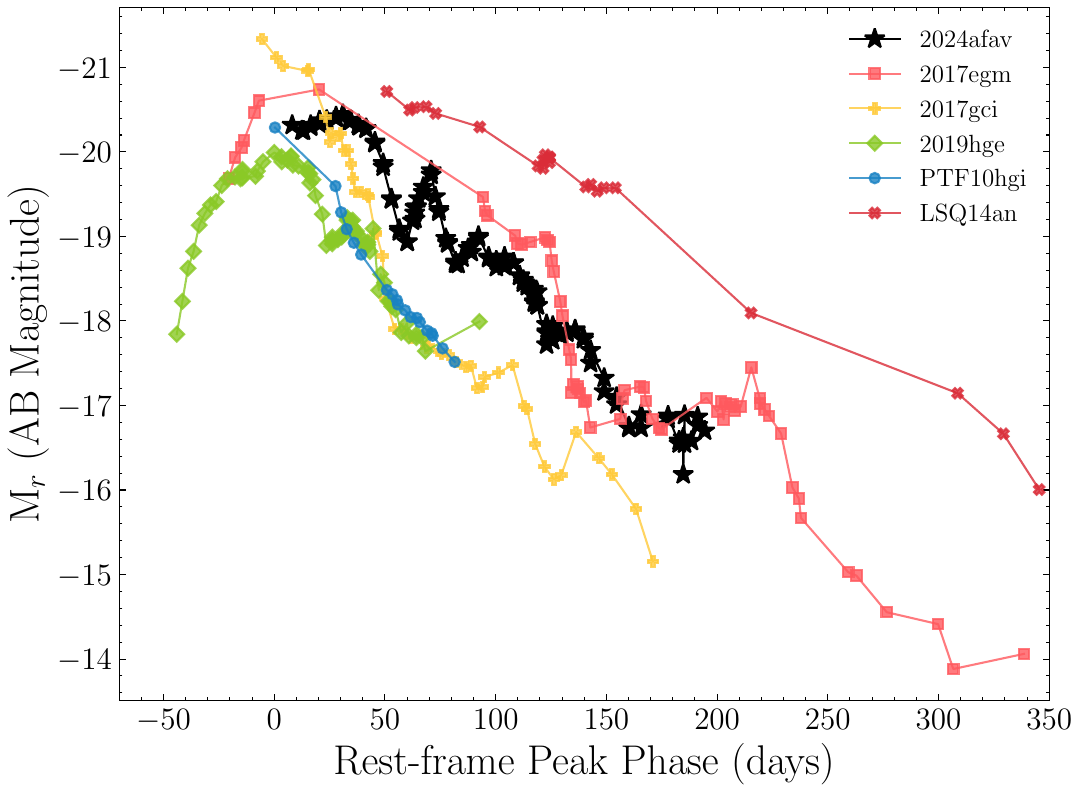}
\caption{The $r$-band light curves of a few SLSNe-I with multiple light curve bumps. All events except PTF10hgi have both photometric and spectral coverage well past $+100$ d, enabling a meaningful comparison. We find that these objects exhibit similar spectral features (see Figure~\ref{fig:compsample}), yet they have dissimilar bump shapes.}
\label{fig:lccomp_a}
\end{figure}

\begin{figure*}[t!]
\centering
\subfigure[]{
    \includegraphics[width=0.45\linewidth]{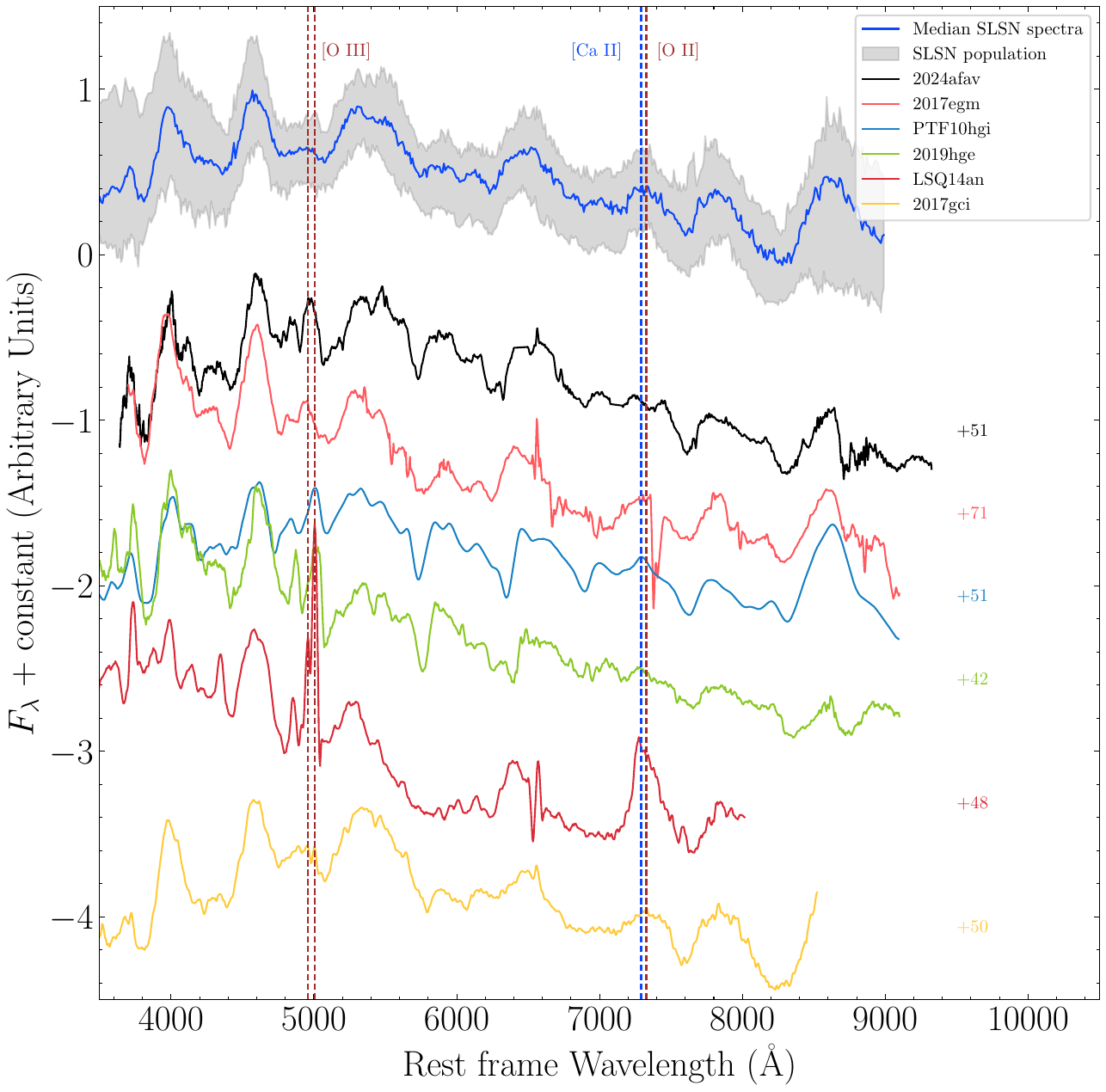}
    \label{fig:speccomp_b}
}
\subfigure[]{
    \includegraphics[width=0.45\linewidth]{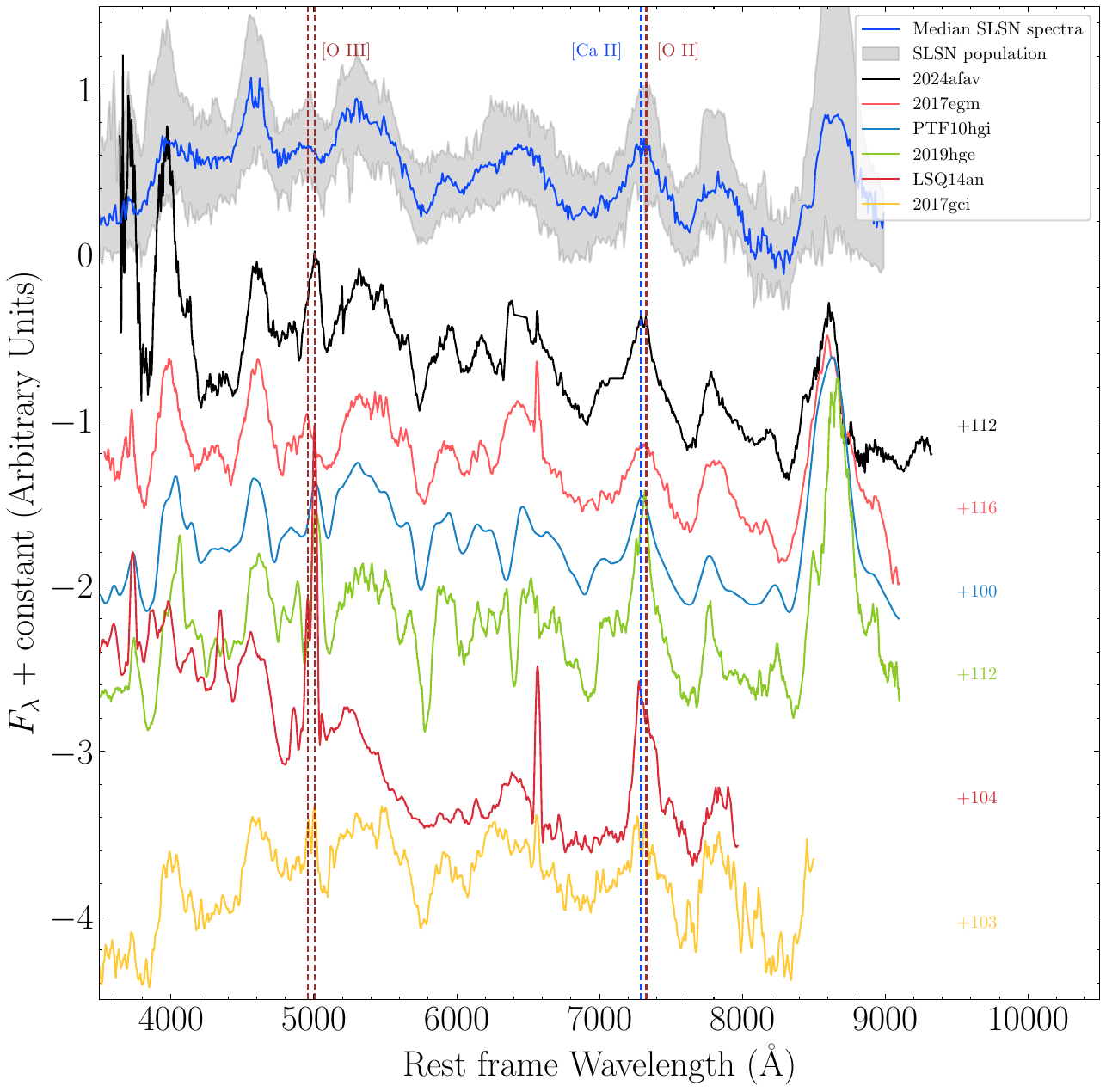}
    \label{fig:speccomp_c}
}
\caption{Comparison of SN\,2024afav spectra at $\approx +50$ d (Left) and $\approx +112$ d with SLSNe-I population from~\citet{2025MNRAS.541.2674A}, and SLSNe-I with multiple bumps \& spectral coverage at similar epochs. These objects represent a small sample of SLSNe-I with multiple bumps in their light curves and a strong presence of the 7300 \AA\, feature. We observe that PTF10hgi, SN\,2019hge, and SN\,2024afav show the presence of [\ion{O}{3}], \ion{He}{1} features at all epochs ,and 7300 \AA\, emission features at $\approx +112$ d phase. The spectral similarity and bumpy light curves of these objects imply a common progenitor. The other two objects in the spectra exhibit bumpy light curves and 7300 \AA\, emission features, indicating CSM interaction and a potential connection between bumps and CSM interaction in SLSNe-I. Both SN\,2017egm, and SN\,2017gci do not show a strong presence of [\ion{O}{3}] features; however, they show [\ion{O}{2}] features at late times, implying weaker ionization in ejecta compared to other objects in the sample.}
\label{fig:compsample}
\end{figure*}

Inspired by the spectral similarity of SN\,2024afav to PTF10hgi and SN\,2017egm, we searched for similar events in the SLSN-I sample~\citep{2024MNRAS.535..471G, 2025arXiv250321874A}, which satisfy the following criteria: (i) the optical light curve exhibits more than two bumps; and (ii) the spectral coverage extends to $\gtrsim +100$ d. We find 4 SLSNe SN\,2017egm, SN\,2017gci, SN\,2019hge and LSQ14an, that satisfy these criteria. We also include PTF10hgi in the comparison sample, despite its limited photometric coverage (until phases of $< +100$ days) and the absence of significant bumps in its light curve, as it exhibits close spectral similarity to SN\,2024afav. We caution that this sample is by no means comprehensive, as the identification is limited to visual inspection with the goal of finding similar events, rather than creating a complete sample. 
The $r$-band light curves SN\,2024afav and the 5 comparison events are shown in Figure~\ref{fig:lccomp_a}. 

We compare the spectra of SN\,2024afav with the 5 comparison events at $\approx +50$ and $+110$ d in Figure~\ref{fig:compsample}. The median SLSN-I spectrum at these phases from \citet{2025MNRAS.541.2674A} is shown for reference. The comparison reveals that 3 of 5 events (SN\,2017egm, SN\,2019hge, and PTF10hgi) exhibit similar spectral features and evolutionary patterns to SN\,2024afav. These events show a remarkably different 5000 \AA\ region compared to median SLSNe-I spectra due to the presence of [\ion{O}{3}] features. In the $\approx +110$-d spectra, these 4 events exhibit a very weak [\ion{O}{1}] line and a stronger 7300 \AA\ feature centered at [\ion{O}{2}]. Another noticeable aspect is the presence of strong absorption at the locations of the \ion{He}{1} lines. In SN\,2024afav (and possibly SN\,2019hge; \citealt{2020ApJ...902L...8Y}), the presence of helium is demonstrated with NIR spectra, In SN\,2017egm and PTF10hgi, previous claims of helium have relied on the optical spectra \citep{2020ApJ...902L...8Y, 2023ApJ...949...23Z}; the similarity to SN\,2024afav confirms these claims. 
We conclude that the spectral and light curve similarities of these 4 events suggest that they share a similar progenitor configuration.

The other two events in the comparison sample, LSQ14an and SN\,2017gci, exhibit different spectral properties. LSQ14an exhibits an early presence of the 7300 \AA\, feature (+48 d), which appears asymmetric and centered on the [\ion{Ca}{2}] lines, and shows no signs of helium features. However, LSQ14an shows a P-Cygni feature at $\approx 6300-6400$ \AA, similar to SN\,2024afav, SN\,2019hge and PTF10hgi, and exhibits strong [\ion{O}{3}]. The presence of strong [\ion{O}{3}] and 7300 \AA\ features, but lack of helium in LSQ14an indicates that its progenitor may have been stripped of its helium layer completely and potentially some of its oxygen layer, contributing to the CSM. SN\,2017gci has a much weaker 7300 \AA\ feature, centered on the [\ion{Ca}{2}] lines, \ion{He}{1} features, and no unusual feature in the $\approx 6300-6400$ \AA\ region. The differences in spectral features between SN\,2024afav and LSQ14an and SN\,2017gci suggests that the light curve bumps do not necessarily require hydrogen or helium in the CSM.

\section{Conclusions} \label{sec:conclusion}

We presented detailed optical/NIR spectroscopic observations of SN\,2024afav, spanning $-14$ to $+160$ d, covering the main peak and multiple post-peak light curve bumps. The key findings are as follows:

\begin{itemize}

    \item SN\,2024afav has an unusual light curve compared to the majority of SLSNe-I, exhibiting seven significant bumps in the post-peak phase.

    \item The pre-peak spectra exhibit typical SLSNe-I features and closely match the early spectra of SN\,2017egm and PTF10hgi.
    \item Post-peak, the spectra show several unusual features indicating CSM interaction with narrow hydrogen and helium shells:
    \begin{itemize}
    \item A narrow, high velocity H$\alpha$ absorption feature, starting at $+20$ d. 
    \item Persistent optical and NIR \ion{He}{1} lines at all available phases. At +23 d, the NIR spectrum showed a high-velocity component of helium at a similar velocity to that of hydrogen, suggesting the presence of helium in both the ejecta and the CSM. 
    \item Early appearance of [\ion{O}{3}] emission starting at $\approx +50$ d. Additionally, a strong [\ion{O}{2}] + [\ion{Ca}{2}] 7300 \AA\ complex starting at $\approx +110$ d.
    \end{itemize}
    

    These spectroscopic features indicate that SN\,2024afav underwent CSM interaction with narrow shells that primarily consist of hydrogen, with a possibly small amount of helium.
    
    \item We compare SN\,2024afav with a subset of SLSNe-I that exhibit bumpy light curves with late time spectral coverage. These objects exhibit some combination of early presence of [\ion{O}{3}], an excess of the [\ion{O}{2}] + [\ion{Ca}{2}] complex, and persistent helium features until late phases, suggestive of CSM interaction. Our comparison sample suggests that the bumps in lightcurves and these unusual features are potentially correlated.
\end{itemize}


In the upcoming era of Rubin Observatory / LSST, detailed late-time light curves of SLSNe-I will be available in abundance. Systematic follow-up of events with multiple bumps will elucidate the fraction of events in which CSM interaction can significantly modify the post-peak light curves and spectra.

\begin{acknowledgments}
We thank Kali Salmas, Alejandra Milone, and  Benjamin Weiner for scheduling the MMT Binospec observations and Yuri Beletsky for performing the Magellan LDSS-3 observations.

The Berger Time-Domain research group at Harvard is supported by the NSF and NASA grants. The LCO supernova group is supported by NSF grants AST-1911151 and AST-1911225. This work is supported by the National Science Foundation under Cooperative Agreement PHY-2019786 (The NSF AI Institute for Artificial Intelligence and Fundamental Interactions, http://iaifi.org/)

KAB is supported by an LSST-DA Catalyst Fellowship; this publication was thus made possible through the support of Grant 62192 from the John Templeton Foundation to LSST-DA.

This paper includes data gathered with the 6.5-meter Magellan Telescopes located at Las Campanas Observatory, Chile. The LCO team is supported by NSF grants AST-2308113 and AST-1911151.

Observations reported here were obtained at the MMT Observatory, a joint facility of the Smithsonian Institution and the University of Arizona. This paper uses data products produced by the OIR Telescope Data Center, supported by the Smithsonian Astrophysical Observatory.

This work makes use of observations from the Las Cumbres Observatory global telescope network. The authors wish to recognize and acknowledge the very significant cultural role and reverence that the summit of Haleakalā has always had within the indigenous Hawaiian community. We are most fortunate to have the opportunity to conduct observations from the mountain. 

We acknowledge the use of public data from the Swift data archive.

This research made use of \sw{PypeIt},\footnote{\url{https://pypeit.readthedocs.io/en/latest/}}
a \sw{python} package for semi-automated reduction of astronomical slit-based spectroscopy
\citep{pypeit:joss_pub, pypeit:zenodo}. This research made use of WISeREP\footnote{\url{ URL https://wiserep.org }}~\citep{2012PASP..124..668Y}.

This work has made use of data from the Zwicky Transient Facility (ZTF). ZTF is supported by NSF grant No. AST- 1440341 and a collaboration including Caltech, IPAC, the Weizmann Institute for Science, the Oskar Klein Center at Stockholm University, the University of Maryland, the University of Washington, Deutsches Elektronen-Synchrotron and Humboldt University, Los Alamos National Laboratories, the TANGO Consortium of Taiwan, the University of Wisconsin–Milwaukee, and Lawrence Berkeley National Laboratories. Operations are conducted by COO, IPAC, and UW. The ZTF forced-photometry service was funded under the Heising-Simons Foundation grant No. 12540303 (PI: Graham).

This work has made use of data from the Asteroid Terrestrial-impact Last Alert System (ATLAS) project. The Asteroid Terrestrial-impact Last Alert System (ATLAS) project is primarily funded to search for near-earth asteroids through NASA grants NN12AR55G, 80NSSC18K0284, and 80NSSC18K1575; byproducts of the NEO search include images and catalogs from the survey area. This work was partially funded by Kepler/K2 grant J1944/80NSSC19K0112 and HST GO-15889, and STFC grants ST/T000198/1 and ST/S006109/1. The ATLAS science products have been made possible through the contributions of the University of Hawaii Institute for Astronomy, the Queen's University Belfast, the Space Telescope Science Institute, the South African Astronomical Observatory, and The Millennium Institute of Astrophysics (MAS), Chile.

This research has made use of the NASA Astrophysics Data System (ADS), the NASA/IPAC Extragalactic Database (NED), and NASA/IPAC Infrared Science Archive (IRSA, which is funded by NASA and operated by the California Institute of Technology) and IRAF (which is distributed by the National Optical Astronomy Observatory, NOAO, operated by the Association of Universities for Research in Astronomy, AURA, Inc., under cooperative agreement with the NSF).

TNS is supported by funding from the Weizmann Institute of Science, as well as grants from the Israeli Institute for Advanced Studies and the European Union via ERC grant No. 725161.

\end{acknowledgments}
\begin{contribution}

\end{contribution}

\vspace{5mm}
\facilities{ATLAS, FLWO 1.2m, LCO, Magellan, MMT, Swift(UVOT), and ZTF}

\software{astropy~\citep{2013A&A...558A..33A,2018AJ....156..123A, 2022ApJ...935..167A}, \sw{SExtractor}~\citep{1996A&AS..117..393B} \sw{NumPy}~\citep{harris2020array}, \sw{photutils}~\citep{2022zndo...6825092B}, \sw{PyRAF}~\citep{2012ascl.soft07011S}, \sw{SciPy}~\citep{2020SciPy-NMeth}, \sw{UVOTSOURCE}, \sw{extrabol}~\citep{2024RNAAS...8...48T}, \sw{PypeIt}~\citep{pypeit:joss_pub}, \sw{lcogtsnpipe}~\citep{2016MNRAS.459.3939V}, and \sw{MOSFiT}~\citep{2017ascl.soft10006G},  }
\newpage

\appendix
\section{Photometry}

\begin{longtable}{|c|c|c|c|} 
\hline
\hline
MJD & Filter & Magnitude $\pm$  e\_magnitude & Telescope \\
\hline
\hline
60665.33	&$	o	$&	18.53	$\pm$	0.09	&	ATLAS	\\
60666.34	&$	o	$&	18.57	$\pm$	0.09	&	ATLAS	\\
60672.65	&$	o	$&	18.03	$\pm$	0.05	&	ATLAS	\\
60674.33	&$	o	$&	17.92	$\pm$	0.04	&	ATLAS	\\
60676.06	&$	c	$&	17.70	$\pm$	0.03	&	ATLAS	\\
60677.31	&$	o	$&	17.85	$\pm$	0.04	&	ATLAS	\\
60678.31	&$	o	$&	17.71	$\pm$	0.03	&	ATLAS	\\
60679.05	&$	c	$&	17.56	$\pm$	0.02	&	ATLAS	\\
60680.04	&$	c	$&	17.54	$\pm$	0.03	&	ATLAS	\\
60681.32	&$	o	$&	17.55	$\pm$	0.03	&	ATLAS	\\
60682.31	&$	o	$&	17.58	$\pm$	0.02	&	ATLAS	\\
60683.06	&$	c	$&	17.36	$\pm$	0.02	&	ATLAS	\\
60684.06	&$	c	$&	17.32	$\pm$	0.02	&	ATLAS	\\
60685.32	&$	o	$&	17.37	$\pm$	0.02	&	ATLAS	\\
60686.30	&$	o	$&	17.40	$\pm$	0.02	&	ATLAS	\\
60687.09	&$	o	$&	17.46	$\pm$	0.03	&	ATLAS	\\
60689.31	&$	o	$&	17.22	$\pm$	0.03	&	ATLAS	\\
60690.49	&$	o	$&	17.20	$\pm$	0.02	&	ATLAS	\\
60691.04	&$	o	$&	17.17	$\pm$	0.03	&	ATLAS	\\
60692.07	&$	o	$&	17.17	$\pm$	0.03	&	ATLAS	\\
60693.22	&$	o	$&	17.13	$\pm$	0.03	&	ATLAS	\\
60697.29	&$	o	$&	17.04	$\pm$	0.03	&	ATLAS	\\
60699.06	&$	o	$&	17.02	$\pm$	0.02	&	ATLAS	\\
60700.05	&$	o	$&	17.06	$\pm$	0.02	&	ATLAS	\\
60701.32	&$	c	$&	16.88	$\pm$	0.01	&	ATLAS	\\
60702.30	&$	c	$&	16.86	$\pm$	0.01	&	ATLAS	\\
60705.30	&$	c	$&	16.9	$\pm$	0.01	&	ATLAS	\\
60706.28	&$	c	$&	16.91	$\pm$	0.01	&	ATLAS	\\
60707.03	&$	o	$&	17.04	$\pm$	0.02	&	ATLAS	\\
60708.02	&$	o	$&	17.11	$\pm$	0.02	&	ATLAS	\\
60709.29	&$	c	$&	16.98	$\pm$	0.02	&	ATLAS	\\
60709.61	&$	UVW1	$&	18.17	$\pm$	0.08	&	Swift	\\
60709.62	&$	U	$&	17.35	$\pm$	0.07	&	Swift	\\
60709.62	&$	B	$&	16.84	$\pm$	0.07	&	Swift	\\
60709.62	&$	UVW2	$&	19.08	$\pm$	0.09	&	Swift	\\
60709.62	&$	V	$&	17.14	$\pm$	0.14	&	Swift	\\
60709.63	&$	UVM2	$&	18.78	$\pm$	0.09	&	Swift	\\
60710.30	&$	c	$&	17.04	$\pm$	0.02	&	ATLAS	\\
60711.01	&$	o	$&	17.12	$\pm$	0.02	&	ATLAS	\\
60711.99	&$	g	$&	16.98	$\pm$	0.01	&	LCO	\\
60711.99	&$	r	$&	17.11	$\pm$	0.01	&	LCO	\\
60711.99	&$	i	$&	17.2	$\pm$	0.02	&	LCO	\\
60712.61	&$	UVW1	$&	18.36	$\pm$	0.14	&	Swift	\\
60712.61	&$	U	$&	17.27	$\pm$	0.12	&	Swift	\\
60712.61	&$	B	$&	16.89	$\pm$	0.13	&	Swift	\\
60712.61	&$	UVW2	$&	19.29	$\pm$	0.16	&	Swift	\\
60712.62	&$	V	$&	16.91	$\pm$	0.22	&	Swift	\\
60712.62	&$	UVM2	$&	18.88	$\pm$	0.14	&	Swift	\\
60714.31	&$	c	$&	17.07	$\pm$	0.01	&	ATLAS	\\
60714.99	&$	o	$&	17.16	$\pm$	0.02	&	ATLAS	\\
60715.03	&$	o	$&	17.15	$\pm$	0.02	&	ATLAS	\\
60715.94	&$	UVW1	$&	18.67	$\pm$	0.1	&	Swift	\\
60715.94	&$	U	$&	17.54	$\pm$	0.09	&	Swift	\\
60716.65	&$	g	$&	17.08	$\pm$	0.01	&	LCO	\\
60716.65	&$	r	$&	17.17	$\pm$	0.02	&	LCO	\\
60716.65	&$	i	$&	17.19	$\pm$	0.02	&	LCO	\\
60717.29	&$	o	$&	17.15	$\pm$	0.03	&	ATLAS	\\
60717.48	&$	r	$&	17.18	$\pm$	0.05	&	FLWO	\\
60717.49	&$	g	$&	16.99	$\pm$	0.05	&	FLWO	\\
60717.50	&$	i	$&	17.27	$\pm$	0.06	&	FLWO	\\
60717.51	&$	z	$&	17.50	$\pm$	0.07	&	FLWO	\\
60718.28	&$	o	$&	17.11	$\pm$	0.04	&	ATLAS	\\
60718.35	&$	B	$&	17.06	$\pm$	0.11	&	Swift	\\
60718.35	&$	V	$&	17.11	$\pm$	0.18	&	Swift	\\
60719.09	&$	o	$&	17.13	$\pm$	0.03	&	ATLAS	\\
60720.92	&$	g	$&	17.07	$\pm$	0.02	&	LCO	\\
60720.92	&$	g	$&	17.07	$\pm$	0.02	&	LCO	\\
60720.93	&$	i	$&	17.08	$\pm$	0.03	&	LCO	\\
60720.93	&$	r	$&	17.11	$\pm$	0.02	&	LCO	\\
60720.93	&$	r	$&	17.13	$\pm$	0.02	&	LCO	\\
60720.93	&$	i	$&	17.16	$\pm$	0.03	&	LCO	\\
60725.18	&$	g	$&	16.97	$\pm$	0.02	&	LCO	\\
60725.18	&$	g	$&	17.00	$\pm$	0.02	&	LCO	\\
60725.19	&$	r	$&	17.08	$\pm$	0.02	&	LCO	\\
60725.19	&$	r	$&	17.09	$\pm$	0.02	&	LCO	\\
60725.19	&$	i	$&	17.14	$\pm$	0.03	&	LCO	\\
60725.19	&$	i	$&	17.17	$\pm$	0.02	&	LCO	\\
60725.21	&$	g	$&	17.01	$\pm$	0.01	&	LCO	\\
60725.21	&$	g	$&	17.02	$\pm$	0.02	&	LCO	\\
60725.21	&$	r	$&	17.05	$\pm$	0.02	&	LCO	\\
60725.21	&$	r	$&	17.06	$\pm$	0.02	&	LCO	\\
60725.22	&$	i	$&	17.12	$\pm$	0.02	&	LCO	\\
60725.22	&$	i	$&	17.12	$\pm$	0.02	&	LCO	\\
60725.34	&$	o	$&	17.04	$\pm$	0.03	&	ATLAS	\\
60726.36	&$	o	$&	17.05	$\pm$	0.02	&	ATLAS	\\
60728.72	&$	g	$&	16.98	$\pm$	0.01	&	LCO	\\
60728.73	&$	g	$&	16.98	$\pm$	0.01	&	LCO	\\
60728.73	&$	r	$&	17.05	$\pm$	0.02	&	LCO	\\
60728.73	&$	r	$&	17.05	$\pm$	0.02	&	LCO	\\
60728.73	&$	i	$&	17.08	$\pm$	0.02	&	LCO	\\
60728.73	&$	i	$&	17.10	$\pm$	0.02	&	LCO	\\
60729.36	&$	o	$&	17.05	$\pm$	0.02	&	ATLAS	\\
60730.36	&$	o	$&	17	$\pm$	0.02	&	ATLAS	\\
60732.90	&$	U	$&	17.53	$\pm$	0.08	&	Swift	\\
60732.90	&$	B	$&	16.89	$\pm$	0.08	&	Swift	\\
60732.90	&$	UVW2	$&	19.86	$\pm$	0.19	&	Swift	\\
60733.20	&$	g	$&	16.89	$\pm$	0.01	&	LCO	\\
60733.20	&$	g	$&	16.91	$\pm$	0.01	&	LCO	\\
60733.21	&$	r	$&	17.01	$\pm$	0.02	&	LCO	\\
60733.21	&$	r	$&	17.02	$\pm$	0.02	&	LCO	\\
60733.21	&$	i	$&	17.03	$\pm$	0.02	&	LCO	\\
60733.21	&$	i	$&	17.07	$\pm$	0.02	&	LCO	\\
60733.35	&$	o	$&	17	$\pm$	0.02	&	ATLAS	\\
60734.26	&$	o	$&	16.92	$\pm$	0.02	&	ATLAS	\\
60736.02	&$	UVW1	$&	18.65	$\pm$	0.11	&	Swift	\\
60736.03	&$	B	$&	17.08	$\pm$	0.09	&	Swift	\\
60736.03	&$	UVW2	$&	19.79	$\pm$	0.14	&	Swift	\\
60736.03	&$	V	$&	17.09	$\pm$	0.15	&	Swift	\\
60736.04	&$	UVM2	$&	19.61	$\pm$	0.13	&	Swift	\\
60736.53	&$	g	$&	16.94	$\pm$	0.01	&	LCO	\\
60736.53	&$	g	$&	16.94	$\pm$	0.01	&	LCO	\\
60736.53	&$	r	$&	16.99	$\pm$	0.01	&	LCO	\\
60736.54	&$	r	$&	16.98	$\pm$	0.01	&	LCO	\\
60736.54	&$	i	$&	17.03	$\pm$	0.02	&	LCO	\\
60736.54	&$	i	$&	17.04	$\pm$	0.02	&	LCO	\\
60737.35	&$	o	$&	16.86	$\pm$	0.02	&	ATLAS	\\
60738.30	&$	o	$&	16.96	$\pm$	0.01	&	ATLAS	\\
60739.42	&$	r	$&	17.06	$\pm$	0.07	&	FLWO	\\
60739.43	&$	g	$&	16.94	$\pm$	0.07	&	FLWO	\\
60739.43	&$	i	$&	17.16	$\pm$	0.07	&	FLWO	\\
60739.44	&$	z	$&	17.32	$\pm$	0.08	&	FLWO	\\
60740.04	&$	c	$&	16.99	$\pm$	0.01	&	ATLAS	\\
60740.24	&$	g	$&	17.07	$\pm$	0.02	&	LCO	\\
60740.25	&$	r	$&	17.05	$\pm$	0.03	&	LCO	\\
60740.25	&$	r	$&	17.06	$\pm$	0.02	&	LCO	\\
60740.25	&$	g	$&	17.07	$\pm$	0.02	&	LCO	\\
60740.25	&$	i	$&	17.09	$\pm$	0.02	&	LCO	\\
60740.26	&$	i	$&	17.10	$\pm$	0.02	&	LCO	\\
60740.53	&$	U	$&	17.92	$\pm$	0.1	&	Swift	\\
60740.53	&$	B	$&	17.13	$\pm$	0.09	&	Swift	\\
60740.54	&$	V	$&	17.29	$\pm$	0.16	&	Swift	\\
60741.34	&$	o	$&	16.96	$\pm$	0.02	&	ATLAS	\\
60742.33	&$	o	$&	16.88	$\pm$	0.02	&	ATLAS	\\
60743.07	&$	c	$&	17.06	$\pm$	0.02	&	ATLAS	\\
60743.98	&$	o	$&	17.09	$\pm$	0.02	&	ATLAS	\\
60744.00	&$	o	$&	17.04	$\pm$	0.04	&	ATLAS	\\
60744.06	&$	g	$&	17.10	$\pm$	0.01	&	LCO	\\
60744.06	&$	g	$&	17.11	$\pm$	0.01	&	LCO	\\
60744.07	&$	r	$&	17.10	$\pm$	0.02	&	LCO	\\
60744.07	&$	r	$&	17.10	$\pm$	0.01	&	LCO	\\
60744.07	&$	i	$&	17.14	$\pm$	0.02	&	LCO	\\
60744.07	&$	i	$&	17.14	$\pm$	0.02	&	LCO	\\
60744.38	&$	r	$&	17.13	$\pm$	0.06	&	FLWO	\\
60744.39	&$	g	$&	17.05	$\pm$	0.07	&	FLWO	\\
60744.39	&$	i	$&	17.19	$\pm$	0.06	&	FLWO	\\
60744.40	&$	z	$&	17.36	$\pm$	0.07	&	FLWO	\\
60745.30	&$	o	$&	16.98	$\pm$	0.02	&	ATLAS	\\
60746.22	&$	o	$&	16.93	$\pm$	0.03	&	ATLAS	\\
60747.04	&$	o	$&	17.13	$\pm$	0.04	&	ATLAS	\\
60747.77	&$	r	$&	17.15	$\pm$	0.03	&	LCO	\\
60747.77	&$	r	$&	17.15	$\pm$	0.02	&	LCO	\\
60747.77	&$	g	$&	17.26	$\pm$	0.03	&	LCO	\\
60747.77	&$	g	$&	17.29	$\pm$	0.02	&	LCO	\\
60747.78	&$	i	$&	17.16	$\pm$	0.04	&	LCO	\\
60747.78	&$	i	$&	17.17	$\pm$	0.04	&	LCO	\\
60748.09	&$	UVW1	$&	19.38	$\pm$	0.25	&	Swift	\\
60748.09	&$	U	$&	17.79	$\pm$	0.16	&	Swift	\\
60748.09	&$	B	$&	17.24	$\pm$	0.16	&	Swift	\\
60748.09	&$	UVW2	$&	20.21	$\pm$	0.28	&	Swift	\\
60748.09	&$	V	$&	17.25	$\pm$	0.27	&	Swift	\\
60748.09	&$	UVM2	$&	19.52	$\pm$	0.3	&	Swift	\\
60752.06	&$	r	$&	17.31	$\pm$	0.03	&	LCO	\\
60752.06	&$	r	$&	17.32	$\pm$	0.03	&	LCO	\\
60752.06	&$	g	$&	17.44	$\pm$	0.03	&	LCO	\\
60752.06	&$	g	$&	17.49	$\pm$	0.02	&	LCO	\\
60752.07	&$	i	$&	17.28	$\pm$	0.04	&	LCO	\\
60752.07	&$	i	$&	17.31	$\pm$	0.04	&	LCO	\\
60753.27	&$	o	$&	17.27	$\pm$	0.03	&	ATLAS	\\
60754.27	&$	o	$&	17.23	$\pm$	0.03	&	ATLAS	\\
60756.02	&$	g	$&	17.91	$\pm$	0.02	&	LCO	\\
60756.02	&$	g	$&	17.93	$\pm$	0.02	&	LCO	\\
60756.03	&$	i	$&	17.55	$\pm$	0.02	&	LCO	\\
60756.03	&$	i	$&	17.55	$\pm$	0.02	&	LCO	\\
60756.03	&$	r	$&	17.56	$\pm$	0.02	&	LCO	\\
60756.03	&$	r	$&	17.60	$\pm$	0.02	&	LCO	\\
60757.23	&$	o	$&	17.61	$\pm$	0.03	&	ATLAS	\\
60758.32	&$	o	$&	17.57	$\pm$	0.03	&	ATLAS	\\
60759.97	&$	o	$&	17.95	$\pm$	0.05	&	ATLAS	\\
60760.02	&$	o	$&	17.97	$\pm$	0.05	&	ATLAS	\\
60760.05	&$	r	$&	17.98	$\pm$	0.02	&	LCO	\\
60760.05	&$	g	$&	18.44	$\pm$	0.02	&	LCO	\\
60760.05	&$	g	$&	18.49	$\pm$	0.02	&	LCO	\\
60760.06	&$	i	$&	17.87	$\pm$	0.02	&	LCO	\\
60760.06	&$	i	$&	17.92	$\pm$	0.03	&	LCO	\\
60760.06	&$	r	$&	17.99	$\pm$	0.02	&	LCO	\\
60761.39	&$	c	$&	18.36	$\pm$	0.07	&	ATLAS	\\
60762.23	&$	c	$&	18.41	$\pm$	0.06	&	ATLAS	\\
60762.98	&$	o	$&	18.27	$\pm$	0.06	&	ATLAS	\\
60763.68	&$	o	$&	18.22	$\pm$	0.03	&	ATLAS	\\
60763.86	&$	g	$&	18.97	$\pm$	0.02	&	LCO	\\
60763.86	&$	g	$&	18.98	$\pm$	0.02	&	LCO	\\
60763.87	&$	i	$&	18.21	$\pm$	0.03	&	LCO	\\
60763.87	&$	i	$&	18.23	$\pm$	0.03	&	LCO	\\
60763.87	&$	r	$&	18.34	$\pm$	0.02	&	LCO	\\
60763.87	&$	r	$&	18.37	$\pm$	0.02	&	LCO	\\
60764.02	&$	o	$&	18.22	$\pm$	0.05	&	ATLAS	\\
60765.24	&$	c	$&	18.67	$\pm$	0.05	&	ATLAS	\\
60766.25	&$	c	$&	18.76	$\pm$	0.07	&	ATLAS	\\
60767.69	&$	r	$&	18.49	$\pm$	0.02	&	LCO	\\
60767.69	&$	g	$&	19.08	$\pm$	0.03	&	LCO	\\
60767.69	&$	g	$&	19.11	$\pm$	0.03	&	LCO	\\
60767.70	&$	i	$&	18.30	$\pm$	0.03	&	LCO	\\
60767.70	&$	i	$&	18.37	$\pm$	0.04	&	LCO	\\
60767.70	&$	r	$&	18.48	$\pm$	0.03	&	LCO	\\
60769.27	&$	c	$&	18.54	$\pm$	0.06	&	ATLAS	\\
60770.23	&$	c	$&	18.47	$\pm$	0.06	&	ATLAS	\\
60770.28	&$	r	$&	18.24	$\pm$	0.04	&	FLWO	\\
60770.29	&$	g	$&	18.70	$\pm$	0.05	&	FLWO	\\
60770.30	&$	i	$&	18.16	$\pm$	0.04	&	FLWO	\\
60770.30	&$	z	$&	18.25	$\pm$	0.07	&	FLWO	\\
60771.32	&$	r	$&	18.15	$\pm$	0.05	&	FLWO	\\
60771.33	&$	g	$&	18.52	$\pm$	0.07	&	FLWO	\\
60771.34	&$	i	$&	18.05	$\pm$	0.05	&	FLWO	\\
60771.34	&$	z	$&	18.16	$\pm$	0.08	&	FLWO	\\
60771.46	&$	i	$&	18.03	$\pm$	0.07	&	LCO	\\
60771.46	&$	r	$&	18.08	$\pm$	0.03	&	LCO	\\
60771.46	&$	r	$&	18.22	$\pm$	0.09	&	LCO	\\
60771.46	&$	g	$&	18.61	$\pm$	0.04	&	LCO	\\
60771.46	&$	g	$&	18.65	$\pm$	0.05	&	LCO	\\
60771.47	&$	i	$&	18.14	$\pm$	0.21	&	LCO	\\
60772.20	&$	r	$&	18.08	$\pm$	0.05	&	FLWO	\\
60772.21	&$	g	$&	18.45	$\pm$	0.05	&	FLWO	\\
60772.22	&$	i	$&	18.02	$\pm$	0.04	&	FLWO	\\
60772.23	&$	z	$&	18.13	$\pm$	0.10	&	FLWO	\\
60773.20	&$	r	$&	17.98	$\pm$	0.06	&	FLWO	\\
60773.21	&$	g	$&	18.33	$\pm$	0.08	&	FLWO	\\
60773.22	&$	i	$&	17.94	$\pm$	0.07	&	FLWO	\\
60773.22	&$	z	$&	18.00	$\pm$	0.09	&	FLWO	\\
60774.19	&$	o	$&	17.81	$\pm$	0.05	&	ATLAS	\\
60775.22	&$	g	$&	18.19	$\pm$	0.05	&	FLWO	\\
60775.50	&$	r	$&	17.84	$\pm$	0.03	&	LCO	\\
60775.50	&$	r	$&	17.90	$\pm$	0.03	&	LCO	\\
60775.50	&$	g	$&	18.32	$\pm$	0.03	&	LCO	\\
60775.50	&$	g	$&	18.34	$\pm$	0.03	&	LCO	\\
60775.51	&$	i	$&	17.75	$\pm$	0.03	&	LCO	\\
60775.51	&$	i	$&	17.84	$\pm$	0.03	&	LCO	\\
60779.29	&$	i	$&	17.64	$\pm$	0.05	&	LCO	\\
60779.29	&$	r	$&	17.64	$\pm$	0.04	&	LCO	\\
60779.29	&$	r	$&	17.70	$\pm$	0.04	&	LCO	\\
60779.29	&$	g	$&	18.27	$\pm$	0.03	&	LCO	\\
60779.29	&$	g	$&	18.29	$\pm$	0.04	&	LCO	\\
60779.30	&$	i	$&	17.70	$\pm$	0.05	&	LCO	\\
60780.40	&$	o	$&	17.8	$\pm$	0.05	&	ATLAS	\\
60781.23	&$	o	$&	17.75	$\pm$	0.05	&	ATLAS	\\
60781.32	&$	r	$&	17.95	$\pm$	0.04	&	FLWO	\\
60781.33	&$	g	$&	18.26	$\pm$	0.07	&	FLWO	\\
60781.33	&$	i	$&	17.88	$\pm$	0.04	&	FLWO	\\
60781.34	&$	z	$&	17.98	$\pm$	0.07	&	FLWO	\\
60782.55	&$	o	$&	17.92	$\pm$	0.03	&	ATLAS	\\
60783.01	&$	g	$&	18.53	$\pm$	0.03	&	LCO	\\
60783.01	&$	g	$&	18.56	$\pm$	0.02	&	LCO	\\
60783.02	&$	i	$&	17.90	$\pm$	0.03	&	LCO	\\
60783.02	&$	i	$&	17.94	$\pm$	0.03	&	LCO	\\
60783.02	&$	r	$&	18.10	$\pm$	0.02	&	LCO	\\
60783.02	&$	r	$&	18.13	$\pm$	0.03	&	LCO	\\
60783.77	&$	o	$&	18.12	$\pm$	0.04	&	ATLAS	\\
60785.25	&$	o	$&	18.19	$\pm$	0.08	&	ATLAS	\\
60786.18	&$	o	$&	18.25	$\pm$	0.05	&	ATLAS	\\
60786.35	&$	r	$&	18.45	$\pm$	0.08	&	FLWO	\\
60786.36	&$	g	$&	18.84	$\pm$	0.09	&	FLWO	\\
60786.36	&$	i	$&	18.40	$\pm$	0.08	&	FLWO	\\
60786.37	&$	z	$&	18.34	$\pm$	0.09	&	FLWO	\\
60787.28	&$	r	$&	18.49	$\pm$	0.03	&	LCO	\\
60787.28	&$	g	$&	19.09	$\pm$	0.02	&	LCO	\\
60787.28	&$	g	$&	19.09	$\pm$	0.02	&	LCO	\\
60787.29	&$	i	$&	18.39	$\pm$	0.03	&	LCO	\\
60787.29	&$	i	$&	18.40	$\pm$	0.03	&	LCO	\\
60787.29	&$	r	$&	18.51	$\pm$	0.03	&	LCO	\\
60789.17	&$	o	$&	18.53	$\pm$	0.04	&	ATLAS	\\
60790.16	&$	o	$&	18.52	$\pm$	0.06	&	ATLAS	\\
60790.90	&$	c	$&	19.13	$\pm$	0.06	&	ATLAS	\\
60791.11	&$	g	$&	19.33	$\pm$	0.02	&	LCO	\\
60791.12	&$	i	$&	18.58	$\pm$	0.03	&	LCO	\\
60791.12	&$	i	$&	18.60	$\pm$	0.03	&	LCO	\\
60791.12	&$	r	$&	18.72	$\pm$	0.02	&	LCO	\\
60791.12	&$	r	$&	18.75	$\pm$	0.02	&	LCO	\\
60791.12	&$	g	$&	19.28	$\pm$	0.02	&	LCO	\\
60791.89	&$	c	$&	19.10	$\pm$	0.08	&	ATLAS	\\
60792.34	&$	r	$&	18.76	$\pm$	0.04	&	FLWO	\\
60792.35	&$	g	$&	19.22	$\pm$	0.05	&	FLWO	\\
60792.35	&$	i	$&	18.65	$\pm$	0.04	&	FLWO	\\
60792.36	&$	z	$&	18.70	$\pm$	0.11	&	FLWO	\\
60793.15	&$	o	$&	18.50	$\pm$	0.06	&	ATLAS	\\
60794.15	&$	o	$&	18.20	$\pm$	0.04	&	ATLAS	\\
60794.27	&$	r	$&	18.68	$\pm$	0.03	&	FLWO	\\
60794.29	&$	g	$&	19.11	$\pm$	0.07	&	FLWO	\\
60794.29	&$	i	$&	18.60	$\pm$	0.05	&	FLWO	\\
60794.30	&$	z	$&	18.55	$\pm$	0.08	&	FLWO	\\
60797.15	&$	o	$&	18.32	$\pm$	0.09	&	ATLAS	\\
60797.22	&$	r	$&	18.57	$\pm$	0.04	&	FLWO	\\
60797.23	&$	g	$&	19.01	$\pm$	0.06	&	FLWO	\\
60797.23	&$	i	$&	18.51	$\pm$	0.05	&	FLWO	\\
60797.24	&$	z	$&	18.41	$\pm$	0.08	&	FLWO	\\
60797.40	&$	g	$&	19.16	$\pm$	0.03	&	LCO	\\
60798.15	&$	o	$&	18.26	$\pm$	0.05	&	ATLAS	\\
60798.73	&$	r	$&	18.53	$\pm$	0.03	&	LCO	\\
60798.73	&$	r	$&	18.61	$\pm$	0.03	&	LCO	\\
60798.73	&$	g	$&	19.04	$\pm$	0.03	&	LCO	\\
60798.73	&$	g	$&	19.06	$\pm$	0.03	&	LCO	\\
60798.74	&$	i	$&	18.38	$\pm$	0.03	&	LCO	\\
60798.74	&$	i	$&	18.42	$\pm$	0.03	&	LCO	\\
60798.91	&$	c	$&	18.74	$\pm$	0.06	&	ATLAS	\\
60799.88	&$	o	$&	18.34	$\pm$	0.05	&	ATLAS	\\
60800.81	&$	o	$&	18.34	$\pm$	0.05	&	ATLAS	\\
60802.19	&$	r	$&	18.42	$\pm$	0.04	&	LCO	\\
60802.19	&$	r	$&	18.44	$\pm$	0.04	&	LCO	\\
60802.19	&$	g	$&	18.97	$\pm$	0.05	&	LCO	\\
60802.19	&$	g	$&	19.02	$\pm$	0.05	&	LCO	\\
60802.20	&$	i	$&	18.36	$\pm$	0.04	&	LCO	\\
60802.20	&$	i	$&	18.40	$\pm$	0.04	&	LCO	\\
60802.39	&$	o	$&	18.22	$\pm$	0.05	&	ATLAS	\\
60807.13	&$	r	$&	18.68	$\pm$	0.07	&	LCO	\\
60807.13	&$	g	$&	18.99	$\pm$	0.07	&	LCO	\\
60807.13	&$	g	$&	19.23	$\pm$	0.08	&	LCO	\\
60807.14	&$	i	$&	18.44	$\pm$	0.09	&	LCO	\\
60807.14	&$	i	$&	18.48	$\pm$	0.08	&	LCO	\\
60807.14	&$	r	$&	18.66	$\pm$	0.08	&	LCO	\\
60810.84	&$	r	$&	18.71	$\pm$	0.04	&	LCO	\\
60810.84	&$	r	$&	18.79	$\pm$	0.04	&	LCO	\\
60810.84	&$	g	$&	19.20	$\pm$	0.05	&	LCO	\\
60810.84	&$	g	$&	19.32	$\pm$	0.05	&	LCO	\\
60810.85	&$	i	$&	18.47	$\pm$	0.04	&	LCO	\\
60810.85	&$	i	$&	18.48	$\pm$	0.04	&	LCO	\\
60810.95	&$	o	$&	18.74	$\pm$	0.10	&	ATLAS	\\
60814.81	&$	r	$&	18.78	$\pm$	0.08	&	LCO	\\
60814.81	&$	g	$&	19.24	$\pm$	0.03	&	LCO	\\
60814.81	&$	g	$&	19.24	$\pm$	0.03	&	LCO	\\
60814.82	&$	i	$&	18.60	$\pm$	0.06	&	LCO	\\
60814.82	&$	r	$&	18.67	$\pm$	0.05	&	LCO	\\
60814.82	&$	i	$&	18.70	$\pm$	0.10	&	LCO	\\
60814.89	&$	o	$&	18.47	$\pm$	0.11	&	ATLAS	\\
60815.06	&$	g	$&	19.28	$\pm$	0.02	&	LCO	\\
60815.06	&$	g	$&	19.30	$\pm$	0.02	&	LCO	\\
60815.07	&$	i	$&	18.59	$\pm$	0.03	&	LCO	\\
60815.07	&$	r	$&	18.75	$\pm$	0.02	&	LCO	\\
60815.07	&$	r	$&	18.76	$\pm$	0.02	&	LCO	\\
60815.08	&$	i	$&	18.63	$\pm$	0.03	&	LCO	\\
60817.14	&$	c	$&	19.08	$\pm$	0.08	&	ATLAS	\\
60818.12	&$	c	$&	19.05	$\pm$	0.08	&	ATLAS	\\
60818.73	&$	o	$&	18.56	$\pm$	0.05	&	ATLAS	\\
60819.18	&$	g	$&	19.25	$\pm$	0.02	&	LCO	\\
60819.19	&$	r	$&	18.73	$\pm$	0.02	&	LCO	\\
60819.19	&$	r	$&	18.74	$\pm$	0.02	&	LCO	\\
60819.19	&$	g	$&	19.26	$\pm$	0.02	&	LCO	\\
60819.20	&$	i	$&	18.64	$\pm$	0.03	&	LCO	\\
60819.20	&$	i	$&	18.65	$\pm$	0.02	&	LCO	\\
60819.87	&$	o	$&	18.80	$\pm$	0.07	&	ATLAS	\\
60821.22	&$	c	$&	19.19	$\pm$	0.12	&	ATLAS	\\
60822.12	&$	g	$&	19.39	$\pm$	0.03	&	LCO	\\
60822.12	&$	g	$&	19.39	$\pm$	0.05	&	LCO	\\
60822.13	&$	r	$&	18.89	$\pm$	0.02	&	LCO	\\
60822.13	&$	r	$&	18.89	$\pm$	0.02	&	LCO	\\
60822.14	&$	i	$&	18.76	$\pm$	0.03	&	LCO	\\
60822.14	&$	i	$&	18.77	$\pm$	0.03	&	LCO	\\
60822.86	&$	o	$&	18.75	$\pm$	0.07	&	ATLAS	\\
60823.88	&$	o	$&	18.80	$\pm$	0.10	&	ATLAS	\\
60823.92	&$	g	$&	19.35	$\pm$	0.02	&	LCO	\\
60823.93	&$	r	$&	18.94	$\pm$	0.02	&	LCO	\\
60823.93	&$	r	$&	18.97	$\pm$	0.02	&	LCO	\\
60823.93	&$	g	$&	19.38	$\pm$	0.02	&	LCO	\\
60823.94	&$	i	$&	18.81	$\pm$	0.04	&	LCO	\\
60823.94	&$	i	$&	18.81	$\pm$	0.04	&	LCO	\\
60825.82	&$	g	$&	19.41	$\pm$	0.02	&	LCO	\\
60825.83	&$	r	$&	19.01	$\pm$	0.02	&	LCO	\\
60825.83	&$	g	$&	19.42	$\pm$	0.02	&	LCO	\\
60825.84	&$	i	$&	18.88	$\pm$	0.03	&	LCO	\\
60825.84	&$	r	$&	18.99	$\pm$	0.02	&	LCO	\\
60825.85	&$	i	$&	18.87	$\pm$	0.03	&	LCO	\\
60826.13	&$	c	$&	19.24	$\pm$	0.09	&	ATLAS	\\
60826.84	&$	o	$&	18.90	$\pm$	0.07	&	ATLAS	\\
60826.91	&$	i	$&	18.77	$\pm$	0.09	&	LCO	\\
60827.84	&$	o	$&	18.90	$\pm$	0.07	&	ATLAS	\\
60828.75	&$	g	$&	19.62	$\pm$	0.03	&	LCO	\\
60828.75	&$	g	$&	19.62	$\pm$	0.03	&	LCO	\\
60828.76	&$	i	$&	18.97	$\pm$	0.04	&	LCO	\\
60828.76	&$	r	$&	19.05	$\pm$	0.03	&	LCO	\\
60828.76	&$	r	$&	19.12	$\pm$	0.03	&	LCO	\\
60828.77	&$	i	$&	18.96	$\pm$	0.04	&	LCO	\\
60829.28	&$	i	$&	19.07	$\pm$	0.05	&	LCO	\\
60829.28	&$	i	$&	19.07	$\pm$	0.05	&	LCO	\\
60829.28	&$	r	$&	19.19	$\pm$	0.03	&	LCO	\\
60829.28	&$	r	$&	19.21	$\pm$	0.03	&	LCO	\\
60829.28	&$	g	$&	19.64	$\pm$	0.03	&	LCO	\\
60829.28	&$	g	$&	19.64	$\pm$	0.03	&	LCO	\\
60830.13	&$	o	$&	18.94	$\pm$	0.15	&	ATLAS	\\
60830.58	&$	r	$&	19.08	$\pm$	0.06	&	LCO	\\
60830.58	&$	i	$&	19.19	$\pm$	0.09	&	LCO	\\
60830.58	&$	i	$&	19.23	$\pm$	0.09	&	LCO	\\
60830.58	&$	r	$&	19.24	$\pm$	0.07	&	LCO	\\
60830.58	&$	g	$&	19.79	$\pm$	0.10	&	LCO	\\
60830.58	&$	g	$&	19.98	$\pm$	0.13	&	LCO	\\
60830.72	&$	g	$&	19.64	$\pm$	0.05	&	LCO	\\
60830.72	&$	g	$&	19.65	$\pm$	0.05	&	LCO	\\
60830.73	&$	i	$&	19.07	$\pm$	0.05	&	LCO	\\
60830.73	&$	r	$&	19.24	$\pm$	0.04	&	LCO	\\
60830.73	&$	r	$&	19.24	$\pm$	0.04	&	LCO	\\
60830.74	&$	i	$&	19.14	$\pm$	0.05	&	LCO	\\
60835.17	&$	g	$&	20.22	$\pm$	0.15	&	LCO	\\
60835.18	&$	r	$&	19.46	$\pm$	0.11	&	LCO	\\
60835.18	&$	r	$&	19.70	$\pm$	0.14	&	LCO	\\
60835.18	&$	g	$&	20.03	$\pm$	0.15	&	LCO	\\
60835.19	&$	i	$&	19.08	$\pm$	0.11	&	LCO	\\
60835.19	&$	i	$&	19.12	$\pm$	0.13	&	LCO	\\
60835.33	&$	i	$&	19.30	$\pm$	0.05	&	LCO	\\
60835.33	&$	i	$&	19.35	$\pm$	0.05	&	LCO	\\
60835.33	&$	r	$&	19.55	$\pm$	0.04	&	LCO	\\
60835.33	&$	r	$&	19.57	$\pm$	0.04	&	LCO	\\
60835.33	&$	g	$&	19.98	$\pm$	0.04	&	LCO	\\
60835.33	&$	g	$&	20.05	$\pm$	0.04	&	LCO	\\
60837.10	&$	g	$&	20.08	$\pm$	0.08	&	LCO	\\
60837.11	&$	r	$&	19.59	$\pm$	0.08	&	LCO	\\
60837.11	&$	g	$&	20.44	$\pm$	0.13	&	LCO	\\
60837.12	&$	i	$&	19.30	$\pm$	0.08	&	LCO	\\
60837.12	&$	i	$&	19.33	$\pm$	0.08	&	LCO	\\
60837.12	&$	r	$&	19.64	$\pm$	0.08	&	LCO	\\
60838.17	&$	g	$&	20.07	$\pm$	0.11	&	LCO	\\
60838.17	&$	g	$&	20.07	$\pm$	0.12	&	LCO	\\
60838.18	&$	r	$&	19.48	$\pm$	0.07	&	LCO	\\
60838.18	&$	r	$&	19.65	$\pm$	0.08	&	LCO	\\
60838.19	&$	i	$&	19.32	$\pm$	0.51	&	LCO	\\
60838.19	&$	i	$&	19.34	$\pm$	0.08	&	LCO	\\
60838.86	&$	o	$&	18.82	$\pm$	0.12	&	ATLAS	\\
60841.76	&$	o	$&	19.37	$\pm$	0.11	&	ATLAS	\\
60841.90	&$	g	$&	20.07	$\pm$	0.08	&	LCO	\\
60841.90	&$	g	$&	20.08	$\pm$	0.09	&	LCO	\\
60841.91	&$	i	$&	19.39	$\pm$	0.06	&	LCO	\\
60841.91	&$	r	$&	19.58	$\pm$	0.06	&	LCO	\\
60841.91	&$	r	$&	19.59	$\pm$	0.06	&	LCO	\\
60841.92	&$	i	$&	19.39	$\pm$	0.06	&	LCO	\\
60843.28	&$	o	$&	19.53	$\pm$	0.19	&	ATLAS	\\
60845.00	&$	g	$&	20.10	$\pm$	0.03	&	LCO	\\
60845.01	&$	r	$&	19.54	$\pm$	0.03	&	LCO	\\
60845.01	&$	r	$&	19.56	$\pm$	0.03	&	LCO	\\
60845.02	&$	i	$&	19.33	$\pm$	0.03	&	LCO	\\
60845.02	&$	i	$&	19.37	$\pm$	0.04	&	LCO	\\
60845.03	&$	o	$&	19.50	$\pm$	0.20	&	ATLAS	\\
60846.26	&$	o	$&	19.16	$\pm$	0.10	&	ATLAS	\\
60848.76	&$	o	$&	19.32	$\pm$	0.11	&	ATLAS	\\
60849.00	&$	g	$&	20.03	$\pm$	0.03	&	LCO	\\
60849.00	&$	g	$&	20.04	$\pm$	0.03	&	LCO	\\
60849.01	&$	r	$&	19.53	$\pm$	0.04	&	LCO	\\
60849.01	&$	r	$&	19.53	$\pm$	0.04	&	LCO	\\
60849.02	&$	i	$&	19.25	$\pm$	0.05	&	LCO	\\
60849.02	&$	i	$&	19.47	$\pm$	0.07	&	LCO	\\
60853.01	&$	g	$&	20.06	$\pm$	0.02	&	LCO	\\
60853.02	&$	r	$&	19.61	$\pm$	0.03	&	LCO	\\
60853.02	&$	r	$&	19.64	$\pm$	0.03	&	LCO	\\
60853.02	&$	g	$&	20.02	$\pm$	0.02	&	LCO	\\
60853.03	&$	i	$&	19.35	$\pm$	0.04	&	LCO	\\
60853.03	&$	i	$&	19.43	$\pm$	0.05	&	LCO	\\
60853.06	&$	o	$&	19.43	$\pm$	0.13	&	ATLAS	\\
60855.80	&$	c	$&	20.08	$\pm$	0.18	&	ATLAS	\\
60856.50	&$	g	$&	20.21	$\pm$	0.06	&	LCO	\\
60856.50	&$	g	$&	20.25	$\pm$	0.06	&	LCO	\\
60856.51	&$	i	$&	19.54	$\pm$	0.09	&	LCO	\\
60856.51	&$	r	$&	19.77	$\pm$	0.06	&	LCO	\\
60856.51	&$	r	$&	19.92	$\pm$	0.07	&	LCO	\\
60856.52	&$	i	$&	19.34	$\pm$	0.07	&	LCO	\\
60863.05	&$	g	$&	20.35	$\pm$	0.13	&	LCO	\\
60863.05	&$	g	$&	20.49	$\pm$	0.13	&	LCO	\\
60863.06	&$	r	$&	20.10	$\pm$	0.12	&	LCO	\\
60863.06	&$	r	$&	20.26	$\pm$	0.14	&	LCO	\\
60863.07	&$	i	$&	19.75	$\pm$	0.12	&	LCO	\\
60863.07	&$	i	$&	19.98	$\pm$	0.15	&	LCO	\\
60869.07	&$	g	$&	21.03	$\pm$	0.18	&	LCO	\\
60869.07	&$	g	$&	21.27	$\pm$	0.28	&	LCO	\\
60869.08	&$	r	$&	20.32	$\pm$	0.12	&	LCO	\\
60869.08	&$	r	$&	20.41	$\pm$	0.12	&	LCO	\\
60869.09	&$	i	$&	20.15	$\pm$	0.13	&	LCO	\\
60869.09	&$	i	$&	20.18	$\pm$	0.13	&	LCO	\\
60875.05	&$	g	$&	20.95	$\pm$	0.06	&	LCO	\\
60875.05	&$	g	$&	20.96	$\pm$	0.07	&	LCO	\\
60875.06	&$	r	$&	20.67	$\pm$	0.08	&	LCO	\\
60875.06	&$	r	$&	20.69	$\pm$	0.08	&	LCO	\\
60875.07	&$	i	$&	20.19	$\pm$	0.09	&	LCO	\\
60875.07	&$	i	$&	20.29	$\pm$	0.10	&	LCO	\\
60880.96	&$	g	$&	20.87	$\pm$	0.14	&	LCO	\\
60880.96	&$	g	$&	21.26	$\pm$	0.14	&	LCO	\\
60880.97	&$	r	$&	20.53	$\pm$	0.09	&	LCO	\\
60880.97	&$	r	$&	20.69	$\pm$	0.11	&	LCO	\\
60880.98	&$	i	$&	20.26	$\pm$	0.11	&	LCO	\\
60880.98	&$	i	$&	20.42	$\pm$	0.11	&	LCO	\\
60894.00	&$	g	$&	20.85	$\pm$	0.19	&	LCO	\\
60894.01	&$	r	$&	20.58	$\pm$	0.17	&	LCO	\\
60894.01	&$	g	$&	20.97	$\pm$	0.22	&	LCO	\\
60894.02	&$	i	$&	20.48	$\pm$	0.19	&	LCO	\\
60894.02	&$	r	$&	20.55	$\pm$	0.16	&	LCO	\\
60894.03	&$	i	$&	20.61	$\pm$	0.20	&	LCO	\\
60899.39	&$	r	$&	20.87	$\pm$	0.11	&	LCO	\\
60899.39	&$	g	$&	21.42	$\pm$	0.11	&	LCO	\\
60899.40	&$	r	$&	20.84	$\pm$	0.08	&	LCO	\\
60899.40	&$	g	$&	21.21	$\pm$	0.10	&	LCO	\\
60901.36	&$	r	$&	20.81	$\pm$	0.20	&	LCO	\\
60901.36	&$	g	$&	21.06	$\pm$	0.19	&	LCO	\\
60901.36	&$	g	$&	21.13	$\pm$	0.35	&	LCO	\\
60901.36	&$	r	$&	21.24	$\pm$	0.47	&	LCO	\\
60901.98	&$	g	$&	21.03	$\pm$	0.08	&	LCO	\\
60901.99	&$	r	$&	20.54	$\pm$	0.08	&	LCO	\\
60901.99	&$	g	$&	21.20	$\pm$	0.09	&	LCO	\\
60902.00	&$	i	$&	20.35	$\pm$	0.13	&	LCO	\\
60902.00	&$	i	$&	20.36	$\pm$	0.10	&	LCO	\\
60902.00	&$	r	$&	20.88	$\pm$	0.11	&	LCO	\\
60905.37	&$	i	$&	20.49	$\pm$	0.16	&	LCO	\\
60905.37	&$	r	$&	20.84	$\pm$	0.19	&	LCO	\\
60905.37	&$	g	$&	21.02	$\pm$	0.16	&	LCO	\\
60908.36	&$	i	$&	20.49	$\pm$	0.21	&	LCO	\\
60908.36	&$	r	$&	20.66	$\pm$	0.24	&	LCO	\\
60908.36	&$	g	$&	20.76	$\pm$	0.23	&	LCO	\\
60908.37	&$	i	$&	20.24	$\pm$	0.14	&	LCO	\\
60908.37	&$	r	$&	20.56	$\pm$	0.21	&	LCO	\\
60908.37	&$	g	$&	21.12	$\pm$	0.28	&	LCO	\\
60911.72	&$	g	$&	21.01	$\pm$	0.25	&	LCO	\\
60911.73	&$	i	$&	19.93	$\pm$	0.28	&	LCO	\\
60911.73	&$	r	$&	20.72	$\pm$	0.23	&	LCO	\\
\hline
\hline
\caption{Photometry of SN\,2024afav. All magnitudes have are in AB system and corrected for Galaxtic extinction.} \label{tab:photometrytable}
\end{longtable}

\begin{longtable}{|c|c|c|c|}
\hline
\hline 
Phase  & \multicolumn{3}{c|}{Velocity (km s$^{-1}$)}\\
 (days)& \ion{Si}{2} $\lambda$6355 \AA&  \ion{O}{1} $\lambda$6158 \AA\, &  \ion{O}{1} $\lambda$7774 \AA\, \\
\hline
 -14 & 11707$\pm$483 &  - & O 6729$\pm$242$^*$ \\
 -4 &  8911$\pm$ 395 & 6540 $\pm$ 54$^*$ & 7037$\pm$ 543 \\
+20 &  7858 $\pm$  41  &   8116 $\pm$ 972   & 7190 $\pm$  141 \\
+24 &  7388  $\pm$  39  &  7473 $\pm$ 49   &  6800 $\pm$ 43  \\
+27 &  7417 $\pm$ 33   &  7387 $\pm$ 46  &  7304 $\pm$  133 \\
+30 &  7171  $\pm$  25  &  6418  $\pm$   95 & 6683 $\pm$  74  \\
+47 &  6282 $\pm$  64   &  6411 $\pm$  126  &  6535 $\pm$  68 \\
+51 &  5443  $\pm$ 43   &  6415 $\pm$  66  & 6341 $\pm$  108 \\
+53 &   -   &  5976 $\pm$ 76$^*$ & 6426  $\pm$  37 \\
+54 &   5439  $\pm$  83  &  6392 $\pm$  105  & 5971 $\pm$ 71  \\
+63 &   4799 $\pm$ 75   &  6663 $\pm$  231  &  5685 $\pm$ 211  \\
+74 &   4786 $\pm$ 84   &  6284  $\pm$  778$^*$ &  4254 $\pm$ 1010$^*$ \\
+91 &   4131 $\pm$  566  &  6355 $\pm$  444  &  5906 $\pm$  127  \\
+112 &   3657 $\pm$  57  &  4879 $\pm$ 162   &  5434 $\pm$  222 \\
+139 &    -    &   5851 $\pm$ 140$^*$ & 5008  $\pm$  51  \\
+160 &   6140 $\pm$  432$^*$ &  5224 $\pm$  104 &  6603 $\pm$ 3289$^*$ \\
\hline
\caption{Velocity evolution measurement using spectral sequence of SN\,2024afav}\label{tab:ph} \\
\end{longtable}

\begin{longtable}{|c|c|c|c|c|c|c|}
\hline
\hline
Phase  & \multicolumn{5}{c|}{Area} & L$_{7300}$/ L$_{6300}$\\
 (days)& [\ion{O}{3}] $\lambda$4959 \AA\, & [\ion{O}{3}] $\lambda$5007 \AA\, & [\ion{O}{1}] $\lambda$6300 \AA\, & [\ion{O}{2}] $\lambda$7319,7330 \AA\,  &  [\ion{Ca}{2}] $\lambda$ 7291,7324 \AA\, & \\
\hline
+112 &  9.29 $\pm$ 2.25  & 48.51 $\pm$ 3.86 & 5.32 $\pm$ 0.57 &  14.26 $\pm$ 3.94  & 11.40 $\pm$ 4.21 & 4.82 $\pm$ 1.14\\
+139 &  10.81 $\pm$ 1.95 &  44.31 $\pm$ 3.16 &  3.12 $\pm$ 0.15 &  65.66 $\pm$ 5.92 &  43.50 $\pm$ 4.87 & 34.99 $\pm$ 2.50 \\
+160 & 4.70 $\pm$  0.92 & 34.72  $\pm$ 1.53 &  6.07 $\pm$ 5.18 & 120.49 $\pm$  5.40 & 206.05  $\pm$ 15.49 & 53.80 $\pm$ 46.24 \\
\hline
\caption{Area n unit of flux density for the the [\ion{O}{3}], [\ion{O}{2}] and [\ion{O}{1}] features, estimated using line profile fitting. We estimated the L$_{7300}$/ L$_{6300}$ ratio for each epoch and found an increasing ratio trend.} \label{tab:featstrength} \\
\end{longtable}

\bibliography{ref}{}
\bibliographystyle{aasjournalv7}

\end{document}